\documentclass[journal ]{new-aiaa}
\usepackage[utf8]{inputenc}
\usepackage{textcomp}

\usepackage{graphicx}
\usepackage{amsmath}
\usepackage[version=4]{mhchem}
\usepackage{siunitx}
\usepackage{longtable,tabularx}
\usepackage{amsmath}
\newcolumntype{P}[1]{>{\centering\arraybackslash}p{#1}}
\usepackage{float}
\usepackage{multirow}
\usepackage{caption}
\usepackage{gensymb}
\usepackage{bm}
\usepackage{booktabs}

\usepackage{lipsum}

\title{Real-Time Adaptive Feedback Control of a Supersonic Dual-Stream Jet\footnotetext{A part of this work was presented as paper 2025-1680 at the AIAA SciTech 2025 Forum, Orlando, FL, January 6-10, 2025.}}

\author{Melissa Yeung
\footnote{Ph.D. Student, Department of Mechanical and Aerospace Engineering; meyeung@syr.edu. Student Member AIAA (Corresponding Author).}
and Yiyang Sun
\footnote{Assistant Professor, Department of Mechanical and Aerospace Engineering. Senior Member AIAA.}
}
\affil{Syracuse University, Syracuse, NY 13244, USA}

\begin{document}

\maketitle

\begin{abstract}
Adaptive control is applied to a supersonic dual-stream jet flow comprised of Mach~1.6 core and Mach~1.0 bypass streams that mix to form a supersonic shear layer. The vortices shed are the source of a high-frequency tone that persists throughout the flow.
The intricate flow dynamics motivates the need for an elaborate and efficient actuation system to suppress the tone and weaken the propagating shock train. 
The present work utilizes online dynamic mode decomposition, which estimates the system dynamics as a locally linear evolution. Snapshot matrices are constructed using sensor measurements, facilitating economical and real-time computations, which are continuously updated and used in a feedback control model. 
Adaptive control is found to efficiently target the resonant tone with little disturbance to the mean features. The framework is not sensitive to sensor placements, enabling actuator design under physically realizable spatial locations in practical implementation. 
To reflect physical limitations, constraints are imposed on the controller model.
It is found that the restricted controller yields greater vortex suppression due to repeated transitory stabilization of the shear layer instability.
Statistical analysis reveals intermittent low-pressure events are responsible for the characteristic frequency, which are largely suppressed by adaptive feedback control.

\end{abstract}

\section*{Nomenclature}

{\renewcommand\arraystretch{1.0}
\noindent\begin{longtable*}{@{}l @{\quad=\quad} l@{}}
$A$                     & area                                          \\
$\bm{A}$                & discrete-time state transition matrix         \\ 
$\mathcal{A}$           & continuous-time state transition matrix       \\ 
$\bm{B}$                & discrete-time control input matrix            \\ 
$\mathcal{B}$           & continuous-time control input matrix          \\
$C_\mu$                 & coefficient of momentum                       \\
$C_T$                   & coefficient of thrust                         \\
$c$                     & speed of sound                                \\ 
$D$                     & computational domain length                   \\
$D_h$                   & hydraulic diameter of rectangular jet exit    \\ 
$F_T$                   & thrust force                                  \\
$f$                     & frequency                                     \\ 
$\bm{K}$                & LQR feedback control gain matrix              \\ 
$M$                     & Mach number                                   \\ 
$m$                     & total number of snapshots                     \\ 
$n$                     & state dimension                               \\ 
$P$                     & pressure                                      \\ 
$P_{xx}$                & autospectral density                          \\
$p$                     & number of control inputs                      \\ 
$q$                     & dynamic pressure                              \\
$R$                     & specific gas constant                         \\ 
$Re$                    & Reynolds number                               \\ 
$St$                    & Strouhal number                               \\
$T$                     & temperature                                   \\ 
$t$                     & time                                          \\ 
$u$                     & streamwise velocity                           \\
$\bm{u}$                & control input                                 \\ 
$W$                     & spanwise splitter plate width                 \\
$w$                     & online DMD window size                        \\ 
$\bm{X}$                & snapshot matrix                               \\ 
$\bm{x}$                & state variable                                \\ 
$\bm{Y}$                & shifted snapshot matrix                       \\ 
$\gamma$                & specific heat ratio                           \\
$\delta_{\text{SP}}$    & splitter plate thickness                      \\
$\delta_{\text{slot}}$  & actuation slot length                         \\
$\delta\tau$            & event temporal extent                         \\
$\Lambda$               & event amplitude                               \\ 
$\mu$                   & dynamic viscosity                             \\ 
$\xi$                   & event definition                              \\ 
$\rho$                  & density                                       \\ 
$\tau$                  & event time                                    \\ 
$\phi$                  & signal reconstruction                         \\
$\psi$                  & angle of actuation                            \\
$\Omega$                & online DMD window weighting factor            \\ 
\end{longtable*}}

\setcounter{table}{0}

\section{Introduction}
\begin{figure}[hbpt]
    \centering
    \includegraphics[width=0.97\textwidth]{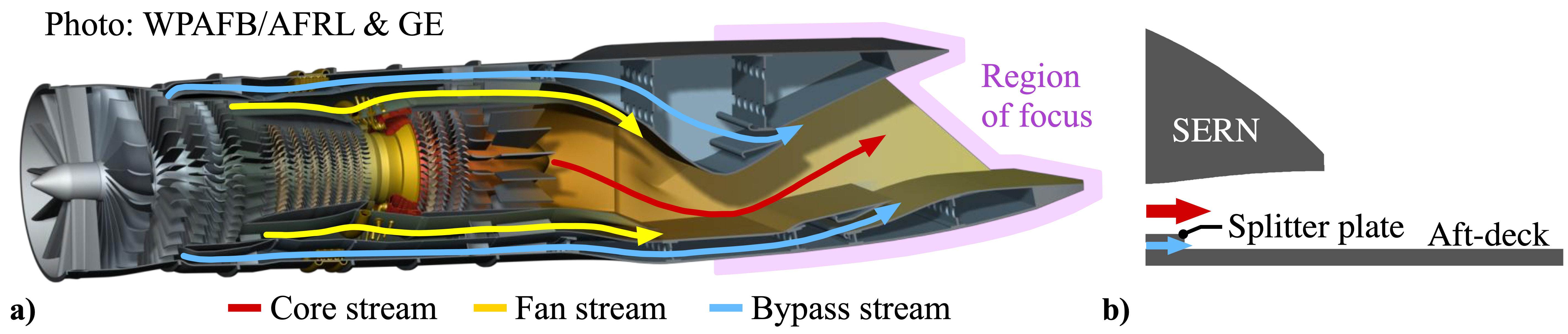}
    \caption{a) Illustration of a generic three-stream engine~\cite{simmons2009design}. b) Nozzle configuration.}
    \label{fig:engine}
\end{figure}

Modern supersonic engines designed for next-generation aircraft have become increasingly complex over the last few decades to meet demanding flight requirements while maintaining propulsive efficiency. 
A variable-cycle engine proposed by Simmons~\cite{simmons2009design}, is shown schematically in Fig.~\ref{fig:engine}a. The arrangement features a rectangular cross-section nozzle to facilitate seamless integration into an airframe.
Additionally, a unique independently modulated cooled bypass stream is designed to continuously adapt to the engine inlet demands while acting as a heat sink to dissipate aircraft heat loads. 

Despite the advancements made in engine design, jet noise remains a persistent problem in the application of jet engines, and is generally associated with turbulent mixing, broadband shock-associated noise, and screech tones. 
Research in jet noise was first pioneered by Lighthill~\cite{lighthill1952sound,lighthill1954sound} and extended to supersonic turbulent shear layers by Phillips~\cite{phillips1960generation}, who suggested the sound field radiates as eddy Mach waves produced along the shear zone. Broadband acoustic waves are additionally generated due to shock cell and shear layer instability interactions. Powell~\cite{powell1953mechanism} further proposed that screech tones are produced when upstream propagating sound waves excite the nozzle lip shear layer, resulting in the closure of a feedback loop. 
Controlling jet noise is critical to aviation safety, for both military and civil operations. 
To this end, various works have sought to improve empirical~\cite{viswanathan2006scaling} and theoretical models~\cite{cabana2008identifying} for jet noise, as well as to identify noise-promoting large-scale structures within turbulent mixing layers~\cite{delville1999examination, ukeiley2001examination, glauser1987coherent}.
For acoustically subsonic jet noise, the fundamental assumption is that the primary contributors to noise sources are composed of intermittent bursts, or events.
Based on this, Kearney-Fischer~\textit{et al.}~\cite{kearney2013intermittent} assumed an extreme limit in which such intermittent bursts are the dominant feature of jet noise, and proposed a method of noise event identification for acoustically subsonic mixing-noise-dominated jets. It was found that a signal reconstruction on these events alone effectively reproduced the important features of the sound-pressure-level spectra at various polar angles. 
For the remainder of this paper, the Mach number being referred to is the local Mach number unless otherwise stated.

In the present work, we analyze and control the supersonic flow at the engine exit, where the core and fan streams are assumed to be fully mixed before entering the nozzle, and are simply referred to as the core stream. 
The nozzle geometry, shown in Fig.~\ref{fig:engine}b, contains a single-sided expansion ramp (SERN), a splitter plate separating the two canonical flows, and an aft-deck plate to mimic the structure of an aircraft. 
This specific engine configuration has been the topic of study over the past decade with joint experimental~\cite{magstadt2017investigating, berry2016investigating, gist2022exploring, kelly2024} and computational efforts~\cite{stack2019turbulence, doshi2023modal, yeung2024two, yeung2024high, thakor2025resolvent}. 
A signature high-frequency tone of approximately 34~kHz is detected in the far-field acoustics, which originates from the instability generated by the mixing of the core and bypass streams. 

Various flow control techniques have been applied with the goal of suppressing the inherent instabilities that dominate this flow. These methods have ranged from passive~\cite{gist2022exploring, doshi2023modal} to active control methods~\cite{kelly2024}. Most notably, Gist~\cite{gist2022exploring} and Doshi~\cite{doshi2023modal} experimentally and computationally introduced a spanwise wavenumber along the trailing edge of the splitter plate, as this region was determined to be most receptive to external forcing~\cite{doshi2023modal, yeung2024high,thakor2025resolvent}. This induced streamwise vorticity that broke down the coherent structures responsible for the high-frequency tone.
The success of this passive control technique prompted an investigation into active flow control techniques~\cite{kelly2024, yeung2024high}, with the ultimate goal of achieving the same control outcomes through an adaptive sensor-actuator control system, to enable the control at a wider range of operating conditions.
Kelly~\cite{kelly2024} experimentally introduced an array of steady-blowing micro-jet actuators along the aft-deck that imitated the troughs of the sinusoidal wave used in Gist's wavy splitter plate. Kelly was able to influence the location of the shock structures, and in some instances split the shock structures in two segments.
Yeung~\textit{et al.}~\cite{yeung2024high} computationally introduced the same micro-jets at various locations on the surfaces of the splitter plate trailing edge and found the strength of the shocks could be influenced by control, consequently attenuating the resultant shock-induced flow separation. While the aforementioned studies detail a comprehensive parametric campaign of steady-blowing actuators and provide significant insights into the actuator location, angle, and mass flow effects, such an approach may not be energy efficient from a practical standpoint and face the risk of failure during off-design flight conditions.

To address this gap, the present work investigates a data-driven approach, where a modal decomposition method is used to build a closed-loop feedback control system. The reduced-order model of the system is continuously (i.e., online) updated using measurements obtained from local probes in the flow.
This approach allows the flow to be driven towards a desirable state with minimal energy input compared to open-loop approaches~\cite{deem2020adaptive}. 
Various data-driven modal decomposition techniques~\cite{taira2017modal} have been historically applied to a multitude of fluid flow problems~\cite{taira2020modal, kalur2018reduced}. These techniques remain an attractive post-processing tool due to their ability to decompose turbulent flow fields into coherent modes that highlight physically important flow features. The present work investigates the use of online dynamic mode decomposition (DMD)~\cite{zhang2019online} in which the DMD algorithm is efficiently computed in real-time and subjected to rank--1 updates, allowing a time-varying state-space model to be generated. This model more accurately reflects the dynamics of the system as it evolves under the influence of external forcing, as opposed to maintaining a time-invariant dynamical model, such as one based on a time-averaged flow state, for feedback control. In these situations, the time-invariant model no longer accurately represents the system once subjected to forcing. Online DMD has been used to suppress a laminar separation bubble using measurements from unsteady surface pressure transducers~\cite{deem2020adaptive}, and holds promise for adaptive control of supersonic flows. In the present study, online DMD is applied to the complex supersonic jet flow problem, which exhibits multiple shear layers, shock-boundary-layer interactions, and vortex-shedding mechanisms.

The computational approach, including the numerical model and flow configuration, the online DMD framework, and the feedback control strategy are presented in Section~\ref{sec:methodsSetup}. Results are shown in Section~\ref{sec:results}, and concluding remarks are given in Section~\ref{sec:conc}.

\section{Numerical Setup and Active Flow Control Methodology}
\label{sec:methodsSetup}

\subsection{Problem Description and Numerical Configuration}
\begin{figure}[hbpt]
    \centering
    \includegraphics[width=0.5\textwidth]{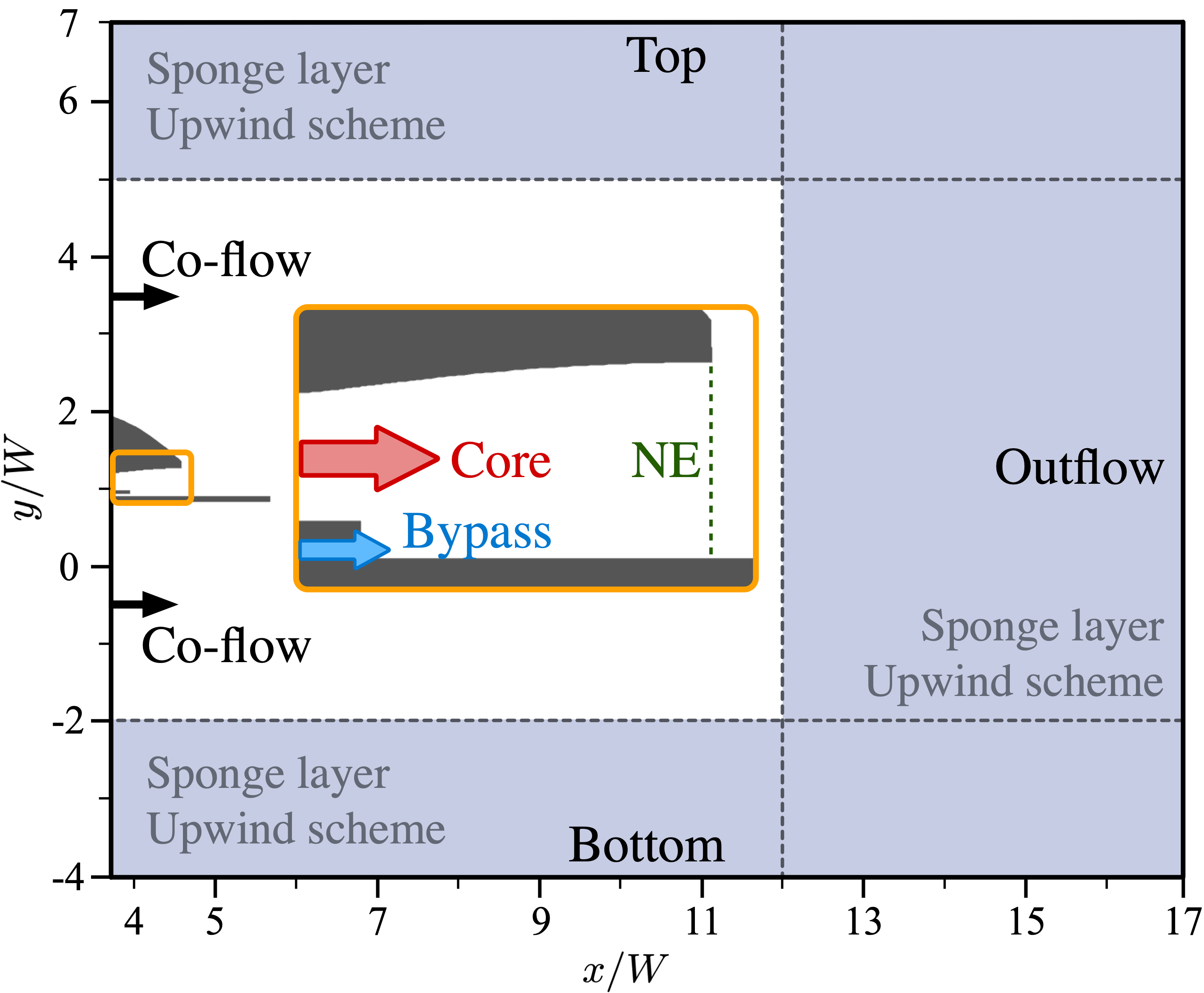}
    \caption{Computational domain and flow configuration with nozzle geometry in gray. NE = Nozzle exit.}
    \label{fig:config_mesh}
\end{figure}

At design conditions, the core stream of the supersonic jet flow operates at a nozzle pressure ratio ($\text{NPR} = P_{\text{total}}/P_{\text{ref}}$) of $\text{NPR}_{\text{core}} = 4.25$, while the bypass stream operates at $\text{NPR}_{\text{bypass}} = 1.89$. This yields Mach numbers of $M_{\text{core}}=1.6$ and $M_{\text{bypass}}=1.0$ for the two streams at the nozzle exit (NE in Fig.~\ref{fig:config_mesh}), based on the isentropic relations.
The nozzle temperature ratio ($\text{NTR} = T_{\text{total}}/T_{\text{ref}}$) is set to unity for both streams and is representative of the unheated jet used in experiments~\cite{berry2016investigating, magstadt2017investigating, kelly2024}. 
The flowfield near the symmetry plane of the nozzle, far from the sidewalls, is approximately two-dimensional (2-D).
Prior results~\cite{yeung2024high} show that 2-D simulations capture the main dynamics of interest and are sufficient for our present goals of developing and demonstrating the feedback control approach with a parametric study.
Simulations are performed using the solver, \textit{CharLES}~\cite{bres2018importance, bres2019modelling}. The solver utilizes a second-order finite-volume method and a third-order Runge-Kutta temporal scheme to solve the compressible Navier-Stokes equations. To capture shocks, flow discontinuities are identified using a hybrid switch~\cite{hill2004hybrid}. Reconstructions are performed with a second-order Essentially Non-Oscillatory method~\cite{shi2002technique}, and the Harten-Lax-van-Leer Contact approximate Riemann~\cite{harten1983upstream} solver computes the flux across a shock. 
The unsteady compressible Navier--Stokes equations are non-dimensionalized using the reference values $\rho_{\text{ref}}=1.173$ kg/m$^3$, $c_{\text{ref}} = 347.189$ m/s, $P_{\text{ref}} = \rho_{\text{ref}} c_{\text{ref}}^2,$ and $T_{\text{ref}} = 300$ K, where $c_{\text{ref}}$ is the reference speed of sound.
The spatial coordinates $x_i$, time $t$, density $\rho$, velocity $u_i$, energy $e$, pressure $P$, and temperature $T$ are non-dimensionalized as
\begin{equation}
    x_i=\frac{x^*}{W^*}, ~~~~~~ t=\frac{t^*c_\text{ref}^*}{W^*}, ~~~~~~ \rho=\frac{\rho^*}{\rho_\text{ref}^*}, ~~~~~~ u_i=\frac{u_i^*}{c_\text{ref}^*}, ~~~~~~ e = \frac{e^*}{\rho_\text{ref}^*(c_\text{ref}^*)^2}, ~~~~~~ P = \frac{P^*}{\rho_\text{ref}^*(c_\text{ref}^*)^2}, ~~~~~~ T = \frac{T^*}{T_\text{ref}^*},
\end{equation}
where the superscript $(\cdot)^*$ refers to the dimensional quantity. $W$ is the splitter plate width in the spanwise direction, which corresponds to an experimental length scale of $W = 82.3$ mm~\cite{berry2016investigating, magstadt2017investigating, kelly2024}.
The Reynolds number is of $Re = \rho_\infty u_\text{jet} D_h/\mu_\infty = 1.5\times10^5$, where $\rho_\infty$ is the freestream density, $u_\text{jet}$ is the nozzle exit velocity calculated based on the isentropic relation and NPR$_\text{core} = 4.25$, $D_h$ is the hydraulic diameter calculated at the nozzle exit, and $\mu_\infty$ is the freestream dynamic viscosity. The hydraulic diameter is calculated to be $D_h = 0.541W$. 

A rectangular domain, shown in Fig.~\ref{fig:config_mesh}, is used to simulate the supersonic flow, where $x$ and $y$ denote the streamwise and wall-normal directions, respectively. 
A structured mesh with non-uniform spacing is used, where the grid around no-slip walls and primary jet flows are refined to resolve the boundary layers and small-scale structures.
Based on a grid resolution study performed in our previous efforts~\cite{yeung2024two, yeung2024high}, a mesh with one million points is sufficient for these simulations.

A co-flow of Mach number $M$ = 0.01 is specified at the left boundary to mimic the experimental setup at the Skytop Turbulence Lab at Syracuse University~\cite{berry2016investigating, magstadt2017investigating, kelly2024}. Characteristic boundary conditions are applied at the core and bypass stream boundaries, where inflow conditions are determined using the designed nozzle pressure and temperature ratios of both streams. 
All the walls of the SERN, splitter plate, and aft-deck (see Fig.~\ref{fig:engine}b) are prescribed to be adiabatic with a no-slip condition. Sponge layers are placed at the top, bottom, and outflow boundaries, where a source term is used to minimize reflection from the boundaries into the domain.
In these regions, a dissipative upwind scheme~\cite{bres2017unstructured} is also used to damp acoustic waves.
The reader is finally referred to Yeung~\textit{et al.}~\cite{yeung2024high} for the comprehensive details on the nozzle geometry, flow configuration, and grid resolution study, where a comparison between the 2-D and 3-D flow features is also given. 

\subsection{Reduced-order Flow Representation using Dynamic Mode Decomposition}
Within the fluid dynamics community, dynamic mode decomposition was first presented and shown to extract insights from high-dimensional data by Schmid~\cite{schmid2010dynamic}. This modal decomposition technique has proved its ability in extracting spatial modes (or DMD modes) from spatio-temporal data, which can be characterized by a frequency and growth rate (or the DMD eigenvalues).
The DMD is a data-driven snapshot-based method, and fundamentally seeks to predict future states of a system by approximating its dynamics as a locally linear evolution. 
Given a linear system $\bm{\dot{x}} = \mathcal{A}\bm{x}$, with $\mathcal{A}$ being the continuous-time dynamics, a discrete-time dynamical system may be constructed as $\bm{x}_{k+1} \approx \bm{Ax}_k$ where a snapshot of the flow $\bm{x}_k$ at some time $t_k$ is related to the next snapshot in time $\bm{x}_{k+1}$ through the linear operator $\bm{A}$. In matrix form, this becomes 
\begin{equation}
    \bm{Y} \approx \bm{AX},
\end{equation}
where $\bm{X}$ and $\bm{Y}$ are 
\begin{equation}
    \bm{X} = \begin{bmatrix}  
                    |       & |        &        & |             \\ 
                   \bm{x}_1 & \bm{x}_2 & \cdots & \bm{x}_{m-1}  \\
                    |       & |        &        & |             \\
                   \end{bmatrix}
    , ~~~~~~~~~~~~
    \bm{Y} = \begin{bmatrix} 
                    |       & |        &        & |             \\ 
                   \bm{x}_2 & \bm{x}_3 & \cdots & \bm{x}_{m}    \\
                    |       & |        &        & |             \\
                  \end{bmatrix}. 
\end{equation}
Here, $\bm{X}$ denotes a sequence of equally-spaced snapshots while $\bm{Y}$ are the snapshots shifted one step into the future, with $m$ being the total number of available snapshots. 
The operator $\bm{A}$ is determined as 
\begin{equation}
    \bm{A} = \bm{YX}^\dagger,
\end{equation}
where $\dagger$ denotes the Moore-Penrose pseudoinverse. The dynamic modes (DMD modes) and associated frequencies and growth rates are then represented through the eigenvectors and eigenvalues of $\bm{A}$, respectively. In practice, when the state dimension $N$ is large, a direct eigendecomposition of $\bm{A}$ may be unfeasible, and it is possible to generate a low-rank representation of $\bm{A}$ through the singular value decomposition (SVD) of $\bm{X}$. DMD can additionally be extended to systems with external forcing, and the reader is referred to the works by Schmid~\cite{schmid2010dynamic, schmid2022dynamic} for further details on the conventional DMD algorithm and its variants.

\subsubsection{Online windowed dynamic mode decomposition}
\label{sec:onlineDMD}
While DMD is a powerful technique to approximate the dynamical system underlying a flowfield, it operates under the assumption that all $m$ snapshots are readily available, and is more commonly used in a post-processing environment. For efficiency, as well as to facilitate control, it becomes essential to consider an approach where the flow representation may be modeled in situ. The present study utilizes an online DMD framework, where the dynamical matrices are not only computed in real-time, but also updated as new snapshots become available~\cite{zhang2019online}. 
The flow dynamics are represented through a linear time-varying system, which can be used to predict future states for control.
Specifically, a windowed variant of online DMD is used, where older snapshots are gradually ``forgotten'' to maintain a constant window size. Assuming the previous snapshot pairs within a finite-time window $w$ are available at some time instant $t_k$, a DMD transition matrix $\bm{A}_k \in \mathbb{R}^{n \times n}$ is found such that $\bm{Y}_k = \bm{A}_k \bm{X}_k$ is satisfied, where 
\begin{equation}
    \bm{X}_k = [\bm{x}_{k-w+1} ~~~~ \bm{x}_{k-w+2} ~~ \cdots ~~ \bm{x}_{k}]
    , ~~~~~~~~~~~~
    \bm{Y}_k = [\bm{y}_{k-w+1} ~~~~ \bm{y}_{k-w+2} ~~ \cdots ~~ \bm{y}_{k}],
\end{equation}
and both $\bm{X}_k$ and $\bm{Y}_k$ are $n \times w$ matrices with $n$ being the state dimension. It is noted that $\bm{X}_k$ and $\bm{Y}_k$ are constructed using sensor data, and thus $n$ also represents the number of sensors being used. The problem then becomes 
\begin{equation}
    \begin{split}
        \bm{A}_k &= \bm{Y}_k \bm{X}_k^\dagger \\
                 &= \bm{Y}_k \bm{X}_k^T(\bm{X}_k \bm{X}_k^T)^{-1} \\ 
                 &= \left[\bm{Y}_k \bm{X}_k^T\right]\left[(\bm{X}_k \bm{X}_k^T)^{-1}\right] \\ 
                 &= \bm{Q}_k \bm{P}_k.
    \end{split}
    \label{eq:oDMDproblem}
\end{equation}
Here, $\bm{X}_k^T(\bm{X}_k \bm{X}_k^T)^{-1}$ is the Moore-Penrose pseudoinverse of $\bm{X}_k$.
It is emphasized that online DMD functions under the assumption that the window size $w$ is at least as large as the state dimension $n$, or $n \leq w$, ensuring $\bm{X}_k \bm{X}_k^T$ can be inverted. The newly defined matrices $\bm{Q}_k$ and $\bm{P}_k$ of size $n \times n$ are updated at each time step $t_{k+1}$ as
\begin{equation}
    \begin{split}
        \bm{Q}_{k+1} &= \bm{Y}_{k+1} \bm{X}_{k+1}^T = \sum_{i = k-w+2}^{k+1} \bm{y}_i \bm{x}_i^T = \bm{Q}_k - \bm{y}_{k-w+1}\bm{x}_{k-w+1}^T + \bm{y}_{k+1}\bm{x}_{k+1}^T , \\ 
        \bm{P}_{k+1}^{-1} &= \bm{X}_{k+1} \bm{X}_{k+1}^T = \sum_{i = k-w+2}^{k+1} \bm{x}_i \bm{x}_i^T = \bm{P}_k^{-1} - \bm{x}_{k-w+1}\bm{x}_{k-w+1}^T + \bm{x}_{k+1}\bm{x}_{k+1}^T . \\
    \end{split}
\end{equation}
This can be interpreted as the $\bm{Q}_k$ and $\bm{P}_k$ matrices forgetting the influence of the oldest snapshot while appending the newest snapshot. 
Letting 
\begin{equation}
    \bm{U} = [\bm{x}_{k-w+1} ~~~ \bm{x}_{k+1}]
    , ~~~~~~~~~~~~
    \bm{V} = [\bm{y}_{k-w+1} ~~~ \bm{y}_{k+1}]
    , ~~~~~~~~~~~~
    \bm{C} = \begin{bmatrix} -1 & 0 \\ 0 & 1 \end{bmatrix},
    \label{eq:UVC}
\end{equation}
$\bm{Q}_{k+1}$ and $\bm{P}_{k+1}$ can be rewritten as 
\begin{equation}
    \bm{Q}_{k+1} = \bm{Q}_k + \bm{VCU}^T
    , ~~~~~~~~~~~~
    \bm{P}_{k+1}^{-1} = \bm{P}_k^{-1} + \bm{UCU}^T.
    \label{eq:updateQ}
\end{equation}
Applying the Sherman-Morrison-Woodbury matrix inversion lemma~\cite{hager1989updating, woodbury1950inverting}, $\bm{P}_{k+1}$ is found to be
\begin{equation}
    \bm{P}_{k+1} = \bm{P}_k - \bm{P}_k \bm{U} \bm{\Gamma}_{k+1} \bm{U}^T \bm{P}_k
    , ~~~~~~~~~~~~
    \bm{\Gamma}_{k+1} = (\bm{C}^{-1} + \bm{U}^T \bm{P}_k \bm{U})^{-1}.
    \label{eq:updateP}
\end{equation}
Substituting Eq. (\ref{eq:updateQ}) and (\ref{eq:updateP}) into (\ref{eq:oDMDproblem}) and rearranging, the update to $\bm{A}$ becomes
\begin{equation}
    \bm{A}_{k+1} = \bm{Q}_{k+1} \bm{P}_{k+1} = \bm{A}_k + (\bm{V} - \bm{A}_k \bm{U})\bm{\Gamma}_{k+1} \bm{U}^T \bm{P}_k.
    \label{eq:updateA}
\end{equation}
Therefore, all updates may be performed using only Eq.~(\ref{eq:UVC}), (\ref{eq:updateP}) and (\ref{eq:updateA}). A real-valued exponential weighting factor $\Omega \in (0,1]$ is additionally incorporated to place higher importance on newer snapshots. 
As $\Omega \to 0$, the algorithm more aggressively attenuates the influence of older snapshots, allowing faster tracking of the system dynamics at the cost of increased sensitivity to noise. 
The choice of $\Omega$ is problem-dependent and should be tailored to the rate at which the system dynamics are changing.
Incorporating into the online DMD algorithm, the update procedure is finalized as
\begin{equation}
    \bm{\tilde{\Gamma}}_{k+1} = (\bm{\tilde{C}}^{-1} + \bm{U}^T \bm{\tilde{P}}_k \bm{U})^{-1}
    , ~~~~~~~~~~~~
    \bm{\tilde{C}} = \begin{bmatrix} -\Omega^w & 0 \\ 0 & 1 \end{bmatrix},
\end{equation}
\begin{equation}
    \bm{A}_{k+1} = \bm{A}_k + (\bm{V} - \bm{A}_k \bm{U})\bm{\tilde{\Gamma}}_{k+1} \bm{U}^T \tilde{\bm{P}}_k, 
\end{equation}
\begin{equation}
    \bm{\tilde{P}}_{k+1} = \frac{1}{\Omega}(\bm{\tilde{P}}_k - \bm{\tilde{P}}_k \bm{U} \bm{\tilde{\Gamma}}_{k+1} \bm{U}^T \bm{\tilde{P}}_k).
\end{equation}

To initialize the algorithm, $\bm{A}_0$ and $\bm{P}_0$ are directly computed using the standard DMD algorithm and the first $w$ snapshots as $\bm{A}_0 = \bm{Y}_w \bm{X}_w^\dag$ and $\bm{P}_0 = (\bm{X}_w \bm{X}_w^T)^{-1}$. In the event that this is unfeasible, $\bm{A}_0$ may be designated to be a zero matrix and $\bm{P}_0 = \alpha \bm{I}$, where $\alpha$ is a large positive scalar value and $\bm{I}$ is the identity matrix~\cite{zhang2019online}. In the present work, the algorithm is initialized with $\bm{A}_0 = \bm{0}$ and $\bm{P}_0 = \alpha \bm{I}$, as direct computation of $\bm{A}_0$ and $\bm{P}_0$ is not easily accomplished in numerical solvers. 

\subsubsection{Online windowed dynamic mode decomposition with control}
From a feedback control standpoint, the control input is also of importance to the dynamical model. The online DMD algorithm is further extended to account for control effects by considering the following discrete-time linear system
\begin{equation}
    \bm{x}_{k+1} = \bm{Ax}_k + \bm{Bu}_k, \label{eq:bmeq}
\end{equation}
where $\bm{x}_k \in \mathbb{R}^{n}$ and $\bm{u}_k \in \mathbb{R}^{p}$, are the states and control inputs, respectively. Here, $p$ denotes the number of control inputs applied to the system. In the context of flow control, $p$ can be interpreted as the number of independent actuators acting on the flow, and each entry of $\bm{u}_k$ represents the control amplitude corresponding to its associated actuator. 
$\bm{A}\in \mathbb{R}^{n \times n}$ and $\bm{B}\in \mathbb{R}^{n \times p}$ are the system and input matrices, respectively, where $\bm{A}$ determines the system evolution through time with no external forcing, and $\bm{B}$ relates the current system state to the control input. 
The same approach used in Section~\ref{sec:onlineDMD} can be used here by simply replacing all the $\bm{X}_k$, $\bm{Y}_k$, and $\bm{A}_k$ terms with
\begin{equation}
    \bm{\tilde{X}}_k = \begin{bmatrix}
                        \bm{x}_{k-w+1} & \bm{x}_{k-w+2} & \cdots & \bm{x}_{k} \\
                        \bm{u}_{k-w+1} & \bm{u}_{k-w+2} & \cdots & \bm{u}_{k} \\
                        \end{bmatrix}
    , ~~~~~~~~~~
    \bm{\tilde{Y}}_k = [\bm{y}_{k-w+1} ~~~~ \bm{y}_{k-w+2} ~~ \cdots ~~ \bm{y}_{k}]
    , ~~~~~~~~~~
    \bm{\tilde{A}}_k = [\bm{A}_k ~~~ \bm{B}_k],
    \label{eq:oDMDc}
\end{equation}
where $\bm{\tilde{X}}_k \in \mathbb{R}^{(n+p) \times w}$, $\bm{\tilde{Y}}_k \in \mathbb{R}^{n \times w}$, and $\bm{\tilde{A}}_k \in \mathbb{R}^{n \times (n+p)}$.
As shown in \ref{sec:apdx_SMD}, online DMD is able to track the dynamics of a time-varying spring-mass-damper system, both with and without the influence of external forcing. 
The reader is finally referred to the work by Zhang~\cite{zhang2019online} for the comprehensive details of the online DMD algorithm and its variants.

\subsection{Closed-Loop Feedback Control System}
\label{sec:feedbackAFC}
Methods from linear control theory may be applied to the approximated discrete-time linear model obtained from online DMD. A feedback controller is adopted using the negative feedback law $\bm{u}_k = -\bm{Kx}_k$, where $\bm{K} \in \mathbb{R}^{p \times n}$. The system becomes 
\begin{equation}
    \bm{x}_{k+1} = (\bm{A} - \bm{BK})\bm{x}_k,
\end{equation}
with the discrete-time state $\bm{A}$ and control $\bm{B}$ matrices being identified by online DMD.
While there have been prior studies investigating various feedback controller designs~\cite{sun2019feedback, yao2022reducing, yao2022feedback}, the present study utilizes a linear quadratic regulator (LQR), and thus $\bm{K}$ is the LQR feedback gain matrix. $\bm{K}$ is determined by minimizing the cost function
\begin{equation}
    J = \sum_{k=0}^\infty (\bm{x}_k^T \bm{Q}_{\text{LQR}} \bm{x}_k + \bm{u}_k^T \bm{R}_{\text{LQR}} \bm{u}_k),
    \label{eq:LQR_cost}
\end{equation}
where the choice of the penalty matrices $\bm{Q}_{\text{LQR}} \in \mathbb{R}^{n \times n}$ and $\bm{R}_{\text{LQR}} \in \mathbb{R}^{p \times p}$ weighs the state and control input performances, respectively. The gain matrix $\bm{K}$ is computed by iterating the discrete-time algebraic Riccati equation and using the stabilizing solution $\bm{D}$ as 
\begin{equation}
    \bm{D}_{i+1} = \bm{A}_k^T \bm{D}_i \bm{A}_k - (\bm{A}^T_k \bm{D}_i \bm{B}_k)(\bm{B}^T_k \bm{D}_i \bm{B}_k + \bm{R}_{\text{LQR}})^{-1} (\bm{A}^T_k \bm{D}_i \bm{B}_k)^T + \bm{Q}_{\text{LQR}},
    \label{eq:DARE}
\end{equation}
\begin{equation}
    \bm{K} = (\bm{R}_{\text{LQR}} + \bm{B}_k^T \bm{D} \bm{B}_k)^{-1} \bm{B}^T_k \bm{D}\bm{A}_k.
    \label{eq:DARE-K}
\end{equation}

The above feedback controller model is used for the supersonic dual-stream jet flow produced by the engine shown in Fig.~\ref{fig:engine}.
Based on the LQR optimization, this feedback control will minimize the measured states $\bm{x}_k$ and the control inputs $\bm{u}_k$. For the entirety of this study, the penalty matrices in Eq.~(\ref{eq:LQR_cost}) are designated to be $\bm{R}_{\text{LQR}} = 1$ and $\bm{Q}_{\text{LQR}} = 20\bm{I}$. 
The measured states are the instantaneous pressure fluctuations, and a total of six sensors are used.

\subsubsection{Active flow control model}

\begin{figure}[hbpt]
\centering
    \includegraphics[width=1\textwidth]{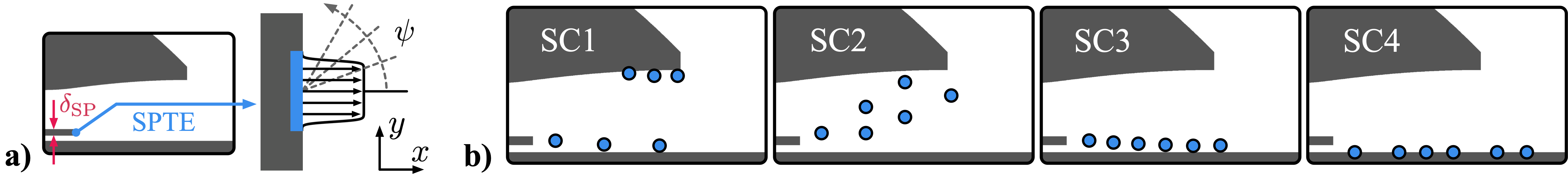}
    \caption{a) AFC schematic with actuation surface indicated in blue. b) Online DMD sensor configurations.}
    \label{fig:AFC_SCs}
\end{figure}

Active flow control (AFC) is introduced in the splitter plate proximity region using micro-jet actuation, as prior input-output analysis has revealed this location to be most sensitive to perturbations~\cite{doshi2023modal, yeung2024high, thakor2025resolvent}. Figure~\ref{fig:AFC_SCs}a depicts a schematic of the AFC configuration, where the actuator is placed along the vertical surface of the splitter plate trailing edge (SPTE). 
The control input $u_k$ is calculated from the negative feedback law $\bm{u}_k = -\bm{Kx}_k$, and is used to define the actuation amplitude, which is bounded such that $|u_k/c_\text{ref}|\leq1.0$ is satisfied unless otherwise noted.
Since the current study employs a single actuator for illustration, $p=1$ and $\bm{u}_k\in\mathbb{R}^p$ reduces to a scalar. Therefore, the scalar notation will be used for the remainder of this paper. 
A top-hat velocity profile $\mathcal{T}(\tilde{y})$ is defined using a hyperbolic tangent function as 
\begin{equation}
    \mathcal{T}(\tilde{y}) = 
        \begin{cases} 
            \dfrac{1}{2}\left[\tanh\left(2000\left(\tilde{y}+\dfrac{\delta_\text{slot}}{2.6}\right)\right)+1\right], & \tilde{y} \leq 0, \\
            \dfrac{1}{2}\left[-\tanh\left(2000\left(\tilde{y}-\dfrac{\delta_\text{slot}}{2.6}\right)\right)+1\right], & \tilde{y} > 0, 
        \end{cases}
\end{equation}
where $\tilde{y} = y-y_\text{c}$, $\delta_\text{slot}$ is the slot length, and $y_c$ denotes the slot center. This profile is used to smoothly distribute the actuation amplitude across the slot and avoid discontinuity at the slot edges. 
The micro-jet is introduced at an angle $\psi$ measured counter-clockwise from the streamwise direction as 
\begin{equation}
    \vec{V}_a =u_k\mathcal{T}(\tilde{y}) \left(\cos(\psi)\vec{i} + \sin(\psi)\vec{j}\right),
\end{equation}
where $\psi = 0\degree$ and $\psi = 30\degree$ are considered, and $\vec{i}$ and $\vec{j}$ denote the streamwise and wall-normal unit vectors, respectively.
$0\degree$ offers the simplest experimental implementation, and previous steady actuation studies considering $\psi \in [-60\degree, 60\degree]$ have shown $30\degree$ to be most beneficial at this location~\cite{yeung2024high}. 
The slot length is designated to be half of the splitter plate thickness as $\delta_\text{slot}=1/2\delta_\text{SP}$, where $\delta_{\text{SP}}$ = 0.0389$W$.

To characterize the actuation, the momentum coefficient $C_\mu$ is defined as
\begin{equation}
    C_\mu = \frac{ \frac{1}{t_\text{tot}} \int_{0}^{t_\text{tot}} \rho_a |\vec{V_a|}^2 A_a ~ dt}{\overline{\rho}_{e,b}~\overline{u}_{e,b}^2 A_e}, 
\end{equation}
where $t_\text{tot}$ is the total time duration of the actuation signal, $\rho_a$ is the density at the actuation slot, $|\vec{V}_a|$ is the local velocity magnitude evaluated at the centerline, and $A_a$ is the actuation area. 
Although $\vec{V}_a$ varies along the slot, the centerline magnitude is used as the representative value for determining $C_\mu$, since the actuation profile is approximately uniform except close to the slot edge, as shown in Fig.~\ref{fig:AFC_SCs}a.
$\overline{\rho}_{e,b}$ and $\overline{u}_{e,b}$, are the time and spatially averaged density and streamwise velocity at the nozzle exit (NE in Fig.~\ref{fig:config_mesh}) for the baseline flow, while $A_e$ is the nozzle exit area.
To characterize the performance of each control case, the thrust coefficient $C_T$ is also calculated as
\begin{equation}
    C_T = \frac{\int_\text{NE} \left[ \overline{\rho}_e(y) ~\overline{u}^2_e(y) +(\overline{P}_e(y)-P_\infty)\right] dy}{P_c A_t},   
    \label{eq:thrust}
\end{equation}
where $\overline{\rho}_e, \overline{u}_e$, and $\overline{P}_e$ are the time-averaged density, streamwise velocity, and pressure at the NE, respectively. The thrust is integrated through the NE and normalized with the chamber pressure $P_c$ and nozzle throat area $A_t$.

To ensure the dynamical model is informed by the appropriate flow dynamics, various sensor configurations (SC) are considered, as shown in Fig.~\ref{fig:AFC_SCs}b. SC1 places sensors along both the shock-induced separation near the nozzle lip and the splitter plate shear layer, and SC2 aims to detect the vortex-induced pressure waves that propagate from the SPTE region. Meanwhile, SC3 distributes sensors within the shear layer, and SC4 restricts all sensors to the aft-deck surface. The sensors in SC4 are also positioned in the same locations as the pressure transducers in the experimental setup~\cite{berry2016investigating}, making SC4 the most practical configuration.

\subsubsection{Physically realizable actuator constraints}

Constraints on the feedback model are considered to better model physical limitations. The sensor data acquisition rate and input signal calculation rate are first both designated to be $St_{D_h} = f D_h / \overline{u}_e = 9.8$, ($\approx~$98~kHz), approximately three times faster than the characteristic frequency of the flow, i.e., the resonant tone frequency. 
This represents an ideal scenario in which the dynamical model is supplied with the temporal resolution required to effectively track the high-frequency shedding associated with the dominant instability.
New control inputs are additionally implemented within the same time instant. 
However, due to the high-speed nature of the supersonic jet flow, the practical application of such a control system is unfeasible. 
Therefore, additional restrictions are imposed on the actuator. 

First, the controller frequency is lowered to $St_{D_h}=0.16$ or 1600~Hz. For the supersonic jet, 1600~Hz represents the lowest possible frequency in which the controller is able to impose new inputs while the flow remains in a transient state. Implementing a frequency lower than 1600~Hz will result in the flow settling to a new steady-state before the controller updates to a new input, and will simply be a string of various steady-control cases. 
Second, the underlying dynamical model estimated by online DMD is allowed to update 10 times before a new input is introduced. This can be interpreted as a time delay between each new input, or the physical limitation associated with the response time of the actuator. 
Finally, subsonic actuation is also considered, where $|u_k/c_\text{ref}|\leq0.3$, as high-frequency forcing can be difficult to achieve with large amplitudes~\cite{cattafesta2011actuators}.
Table~\ref{tab:afc_cases} tabulates all control cases considered in the present study, where SC4 is chosen for the additional constraints. This is due to the minimal differences observed between sensor configurations, as will be discussed in Section~\ref{subsec:adptvFC}. 
\begin{table}[hbpt]
\begin{center}
\begin{tabular}{P{4cm} P{2cm} P{2cm} P{4cm}}
     \toprule
     Sensor configuration & $\psi$ & $|u_k/c_\text{ref}|$ & Controller frequency, $St_{D_h}$ \\ 
     \hline
     SC1, SC2, SC3, SC4 & $0\degree,~30\degree$ & $\leq1.0$         & 9.80 \\ 
     SC4                & $30\degree$           & $\leq1.0,~\leq0.3$ & 0.16 \\ 
     \bottomrule
\end{tabular}
\caption{Summary of the feedback control configurations.}
\label{tab:afc_cases}
\end{center}
\end{table}

\subsection{Event identification and signal reconstruction}
\label{sec:event_method}

\begin{figure}[hbpt]
\centering
    \includegraphics[width=1\textwidth]{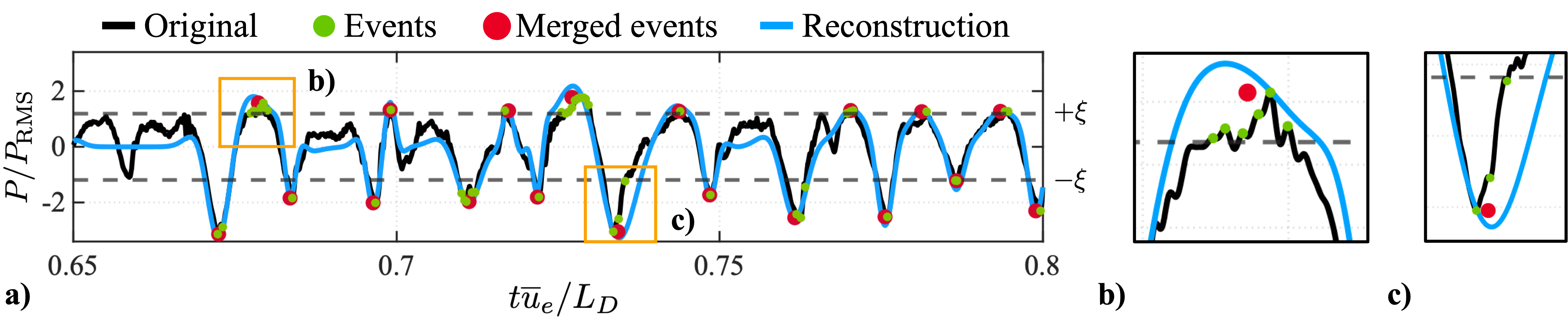}
    \caption{a) Method of event identification and signal reconstruction. b-c) Closer view of events.}
    \label{fig:event_method}
\end{figure}

By assuming the primary contributors to noise sources in an acoustically subsonic and mixing-dominated jet consists of intermittent bursts, or events, a signal can be reconstructed using the time and amplitude of each event, and their associated temporal extent~\cite{kearney2013intermittent}. 
Although this jet flow problem is locally supersonic, i.e., $M=u/\sqrt{\gamma RT}>1$, the flow within the splitter plate shear layer, wherein the dominant frequency originates, is found to be acoustically subsonic, i.e., $M_a=u/\sqrt{\gamma RT_\infty}<1$. 
In this analysis, an event is defined as any instant in which a local signal peak exceeds $\pm~\xi P_\text{RMS}$, where $\xi$ is a scalar value that is unique to the signal and flow condition~\cite{kearney2013intermittent}. 
The choice of $\xi$ will be explored in Section~\ref{sec:events}. 
Figure~\ref{fig:event_method}a demonstrates an example of the signal reconstruction process, where all local events are marked with green dots. 
For each event, the time $\tau_i$, amplitude $\Lambda_i$, and temporal extent $\delta\tau_i$ are identified, where $\delta\tau_i$ is defined as the full width at half-maximum (FWHM). 
This is determined by starting at $\tau_i$ for some event $i$, and searching outwards in both directions until the criterion is achieved. 
Due to the noise present in the original signal, events are often identified in clusters, as shown in Fig.~\ref{fig:event_method}b-c. Therefore, any events with overlapping temporal extents $\delta\tau_i$ are merged and considered to be one event, as indicated with red dots. Each event $i$ is then modeled using the Mexican hat function as
\begin{equation}
    \phi_i(t) = \Lambda_i\left(1-\frac{(t-\tau_i)^2}{(\delta\tau_i \epsilon)^2}\right) \text{exp}\left[\frac{-(t-\tau_i)^2}{(\delta\tau_i\epsilon)^2}\right],
\end{equation}
where $\epsilon$ is an adjustment factor that allows the FWHM to be used as a characteristic scale, and determined to be $\epsilon=0.9$~\cite{kearney2013intermittent}. 
The final reconstruction is achieved through a superposition of all $\phi_i(t)$.
This reconstruction can be interpreted as isolating the intermittent events and discarding all other information from the original signal, allowing investigation into event-focused contributions to the dominant tone. 

\section{Results}
\label{sec:results}

In this section, the baseline flow features of the dual-stream jet are presented. 
Then, adaptive control is implemented using online DMD. The supersonic flow under the influence of adaptive control is compared with open-loop steady-blowing control.
Finally, restrictions on the control method are explored and a statistical analysis on intermittent events is provided.

\subsection{Dual-Stream Jet Flow}

\subsubsection{Baseline flow features}
\begin{figure}[hbpt]
\centering
    \includegraphics[width=0.9\textwidth]{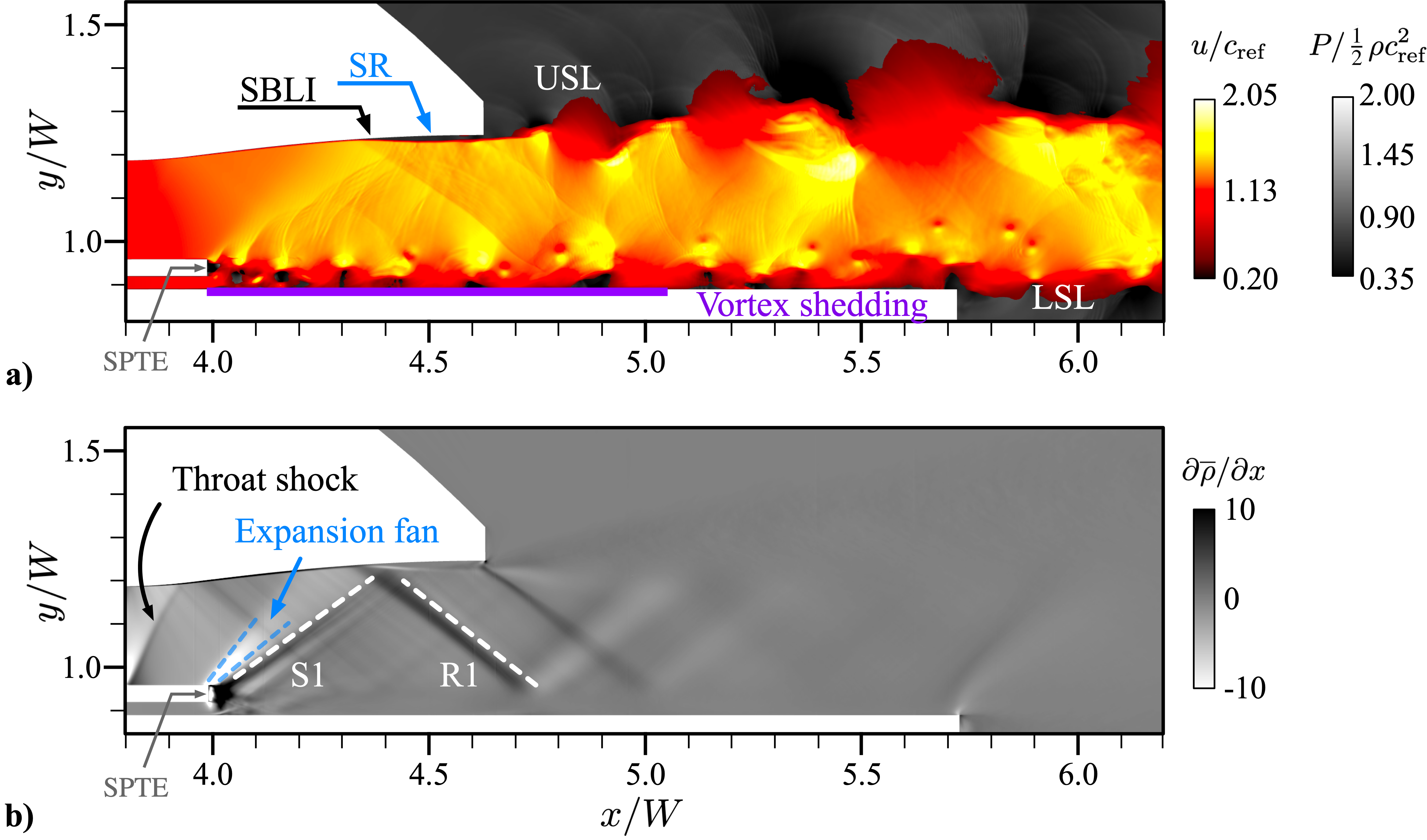}
    \caption{Instantaneous a) and mean b) flow fields for baseline case.}
    \label{fig:base_flow}
\end{figure}

The supersonic dual-stream jet flow is comprised of a variety of features, with several undesired manifestations.
An instantaneous and mean flow field of the baseline case is shown in Fig.~\ref{fig:base_flow}. Figure~\ref{fig:base_flow}a displays shaded color contours of the streamwise velocity together with black-white contours of the pressure field, while \ref{fig:base_flow}b shows the time-averaged density gradient in the $x$-direction with negative and positive values highlighting expansions and shocks, respectively. 
A relatively strong oblique shock system is observed aft of the SPTE, denoted as S1.
S1 impinges on the expansion ramp, resulting in a shock-boundary-layer interaction (SBLI) and consequently a separation region (SR), before reflecting as R1. 
S1 is comprised of an oblique shock and vortex-induced compression waves. 
The former is generated as the core stream aligns with the horizontal direction downstream of the SPTE, while the latter are associated with vortex shedding generated due to the mixing of the core and bypass streams. 
Near the SPTE, the shed vortices impose substantial unsteady loads along the aft-deck surface. 
Upper and lower shear layers (USL \& LSL) form between the bulk flow exiting the nozzle and the ambient. 

\subsubsection{Spectral analysis of flow unsteadiness}

\begin{figure}[hbpt]
\centering
    \includegraphics[width=1\textwidth]{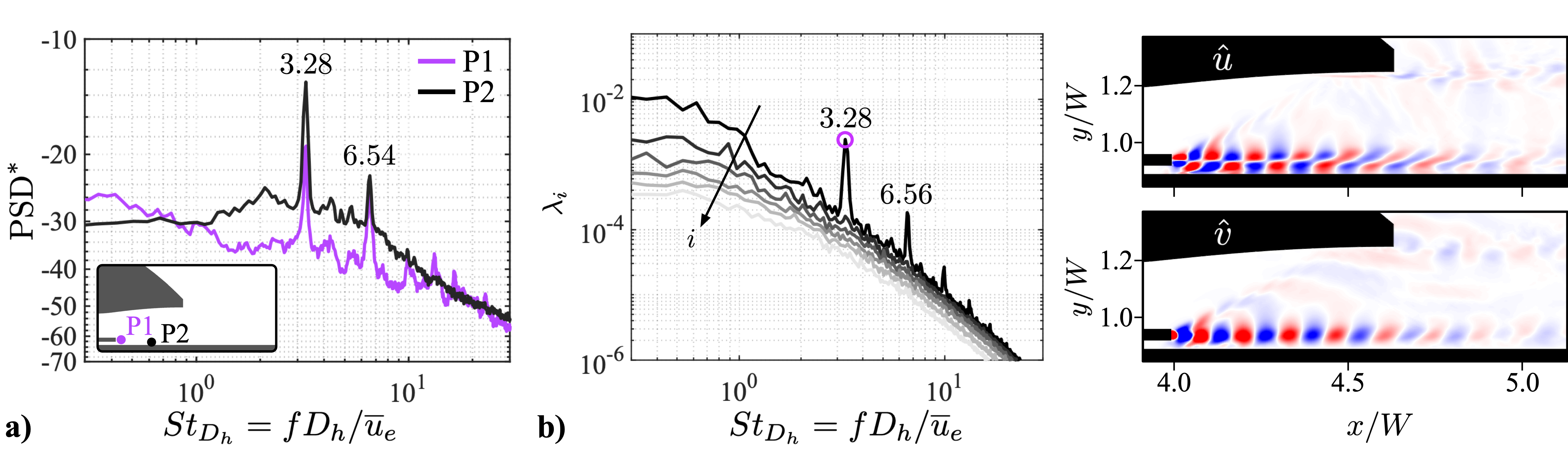}
    \caption{a) PSD$^*$ for the baseline flow. b) SPOD spectra with leading modes at the resonant tone.}
    \label{fig:base_spect}
\end{figure}

The vortex shedding originating from the splitter plate trailing edge gives rise to the characteristic frequency (i.e., resonant tone) of the flow. The frequency spectrum is first examined using the power spectral density (PSD) of the pressure time series at two point probes, denoted P1 and P2 in Fig.~\ref{fig:base_spect}. Figure~\ref{fig:base_spect}a shows the non-dimensional PSD, $\text{PSD}^* = 10\text{log}_{10} (P_\text{xx} u_\text{jet}/q_\infty^2 D_h)$, and the probe locations, where $P_\text{xx}$ is the autospectral density and $q_\infty$ is the free-stream dynamic pressure. The P1 probe is placed within the splitter plate wake $(x,y)/W = (4.017, 0.940)$ and directly captures the dominant shedding frequency of $St_{D_h} = 3.28$ and its harmonics. This frequency is the same tone captured numerically by Stack~\cite{stack2019turbulence} (33~kHz) and experimentally in the far-field acoustics by Berry~\cite{berry2016investigating} (34~kHz) of approximately $St_{D_h} = 3.3$ and highlights the appropriateness of the current simulation approach. Further downstream within the splitter plate shear layer, the resonant tone persists in the P2 probe $(x,y)/W = (4.301, 0.921)$. Since actuation along the SPTE surface interferes with the signal through P1 (see Fig.~\ref{fig:AFC_SCs}a), spectra at P2 will be used for the remainder of the study to explore changes in the dominant tone.

Spectral proper orthogonal decomposition (SPOD)~\cite{glauser1987coherent, towne2018spectral, schmidt2020guide} is applied using the non-dimensional flow variables $\bm{q} = [\rho~u~v~T]'$ and weighted according to Chu's compressible energy norm~\cite{chu1965energy}. 5000 snapshots are collected at equally spaced time-steps $\Delta t \overline{u}_e/L_D=8.29\times10^{-4}$, where $L_D=13.25W$ is the length of the entire computational domain. 
The snapshots are partitioned into 15 blocks with 75\% overlap, such that each block contains 1111 snapshots.
Spatio-temporal coherent structures are extracted from the flow field to gain further insight into the base flow. Figure~\ref{fig:base_spect}b shows the SPOD energy spectra and the leading mode associated with the resonant tone $St_{D_h}=3.28$. It is observed that the cause of the high-energy resonant tone is concentrated within the splitter plate shedding, and this instability largely dominates the baseline flow.

\subsection{Adaptive Control of Jet Flow Using Online DMD}
\label{subsec:adptvFC}

\begin{figure}[hbpt]
\centering
    \includegraphics[width=1\textwidth]{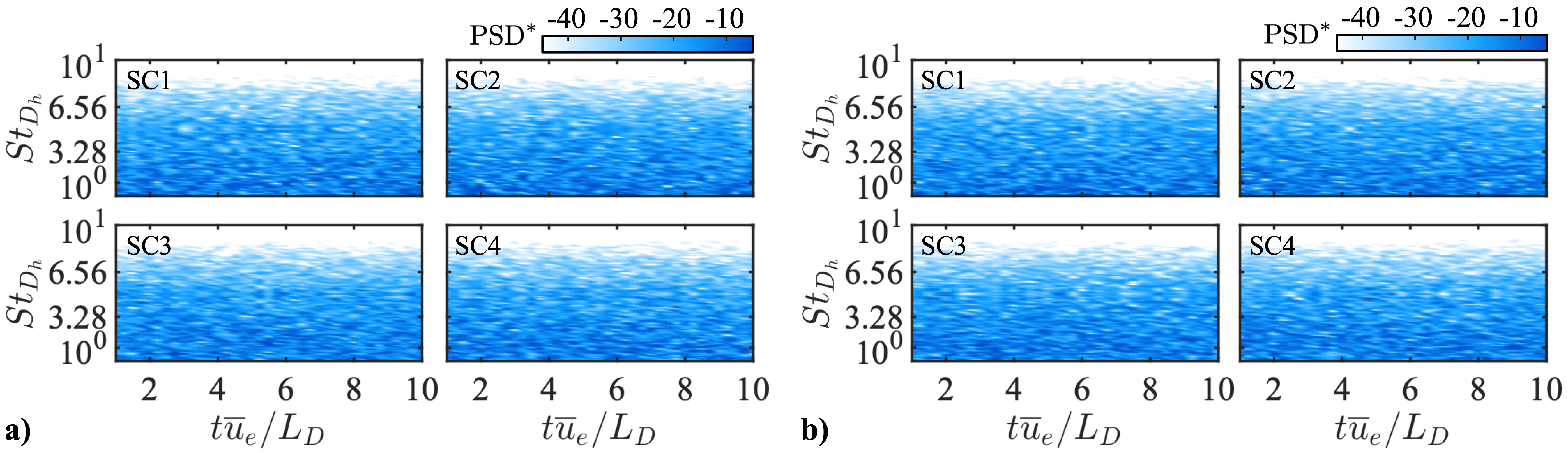}
    \caption{Spectrogram of the actuation signal for adaptive control introduced at a) $\psi=0\degree$ and b) $\psi=30\degree$.}
    \label{fig:ADFC_ukSpect}
\end{figure}

Adaptive control using online DMD is first explored without constraints on actuator frequency or actuator flow speed, with data collection and model update rates being performed at $St_{D_h}=9.8$. 
Actuation is introduced at either $\psi=0\degree$ or $30\degree$, and the PSD$^*$ of the control signal $u_k/c_\text{ref}$, obtained from the procedure of Section~\ref{sec:feedbackAFC}, is shown as a spectrogram for all unconstrained cases in Fig.~\ref{fig:ADFC_ukSpect}a and \ref{fig:ADFC_ukSpect}b, respectively. 
All spectrograms are computed such that each window covers a time period of $t\overline{u}_e/L_D=0.5$ with $50\%$ overlap, and is consistent for the remainder of this paper.
The resonant tone $St_{D_h}=3.28$ and its harmonic $St_{D_h}=6.56$ are also indicated along the $y$-axis for reference.
The control signal is always active in time and does not exhibit a dominant frequency. As shown in the spectrograms, the input contains numerous time-varying frequencies that coexist at each time instant. The active frequencies span a wide range, indicating that adaptive control with online DMD engages the entire available frequency range. 
Rather than favoring a select portion, the spectrogram is mostly broadband. 
Lastly, the control signal appears to have little difference between each sensor configuration and actuation angle, indicating that online DMD may be implemented with more restrictive or practically feasible sensor configurations. 

\subsubsection{Instantaneous and mean flow modifications}

\begin{figure}[hbpt]
\centering
    \includegraphics[width=1\textwidth]{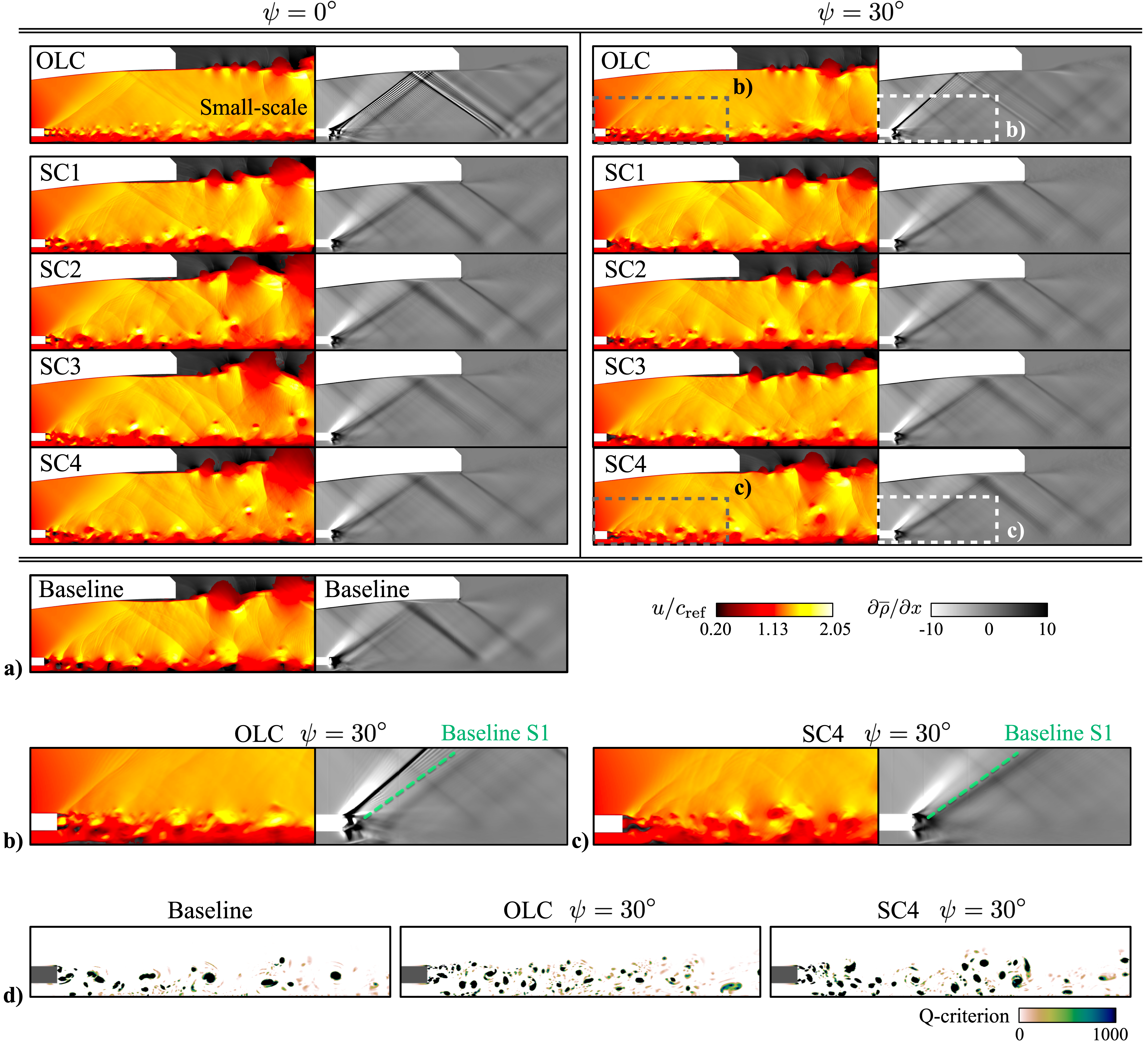}
    \caption{a) Instantaneous and mean flow fields for all adaptive control cases. b-c) Closer view near the SPTE. d) Contours of the Q-criterion.}
    \label{fig:ADFC_flow}
\end{figure}

The instantaneous and mean flow fields for all unconstrained cases are shown in Fig.~\ref{fig:ADFC_flow}a together with the baseline and steady-blowing, open-loop control (OLC) results from our prior study~\cite{yeung2024high} for comparison. 
A closer view of the flow near the SPTE for representative OLC and unconstrained adaptive control cases are given in Fig.~\ref{fig:ADFC_flow}b-d, where Fig.~\ref{fig:ADFC_flow}d illustrates contours of the Q-criterion to highlight vortex cores. 
For both steady OLC and unsteady adaptive control, the actuator directly interferes with the mixing of the main and bypass streams. 
The large coherent structures in the baseline flow are broken down into smaller and less organized structures. 
Steady OLC produces the smallest-scale structures relative to the baseline, while adaptive control allows some medium-sized structures to form intermittently.

Regarding shock patterns for adaptive control, actuating at $\psi = 30\degree$ causes the primary S1 shock to appear slightly weaker compared to $\psi = 0\degree$ for some sensor configurations, however, the mean flows and overall shock structures are largely unaffected across all adaptive control cases. 
Meanwhile, employing steady OLC at the same locations and angles results in a similar breakdown of large-scale structures associated with the shedding instability. However, the steady-blowing cases drastically alter the mean, such that the overall flow has deviated from the nominal configuration. 
As shown in Fig.~\ref{fig:ADFC_flow}b-c, OLC results in a noticeable spatial displacement of S1 relative to the baseline, while adaptive control has minimal influence on S1.
In contrast, adaptive control effectively disrupts the formation of the high-energy structures while preserving the fundamental features of the baseline flow.
This can be particularly advantageous in situations where a baseline configuration is considered mostly optimal for the primary objective, but introduces a problematic secondary consideration. 
In these cases, adaptive control using online DMD can efficiently target an instability without altering the overall flow problem.

\subsubsection{Reduced unsteadiness through adaptive control}

\begin{figure}[hbpt]
\centering
    \includegraphics[width=0.8\textwidth]{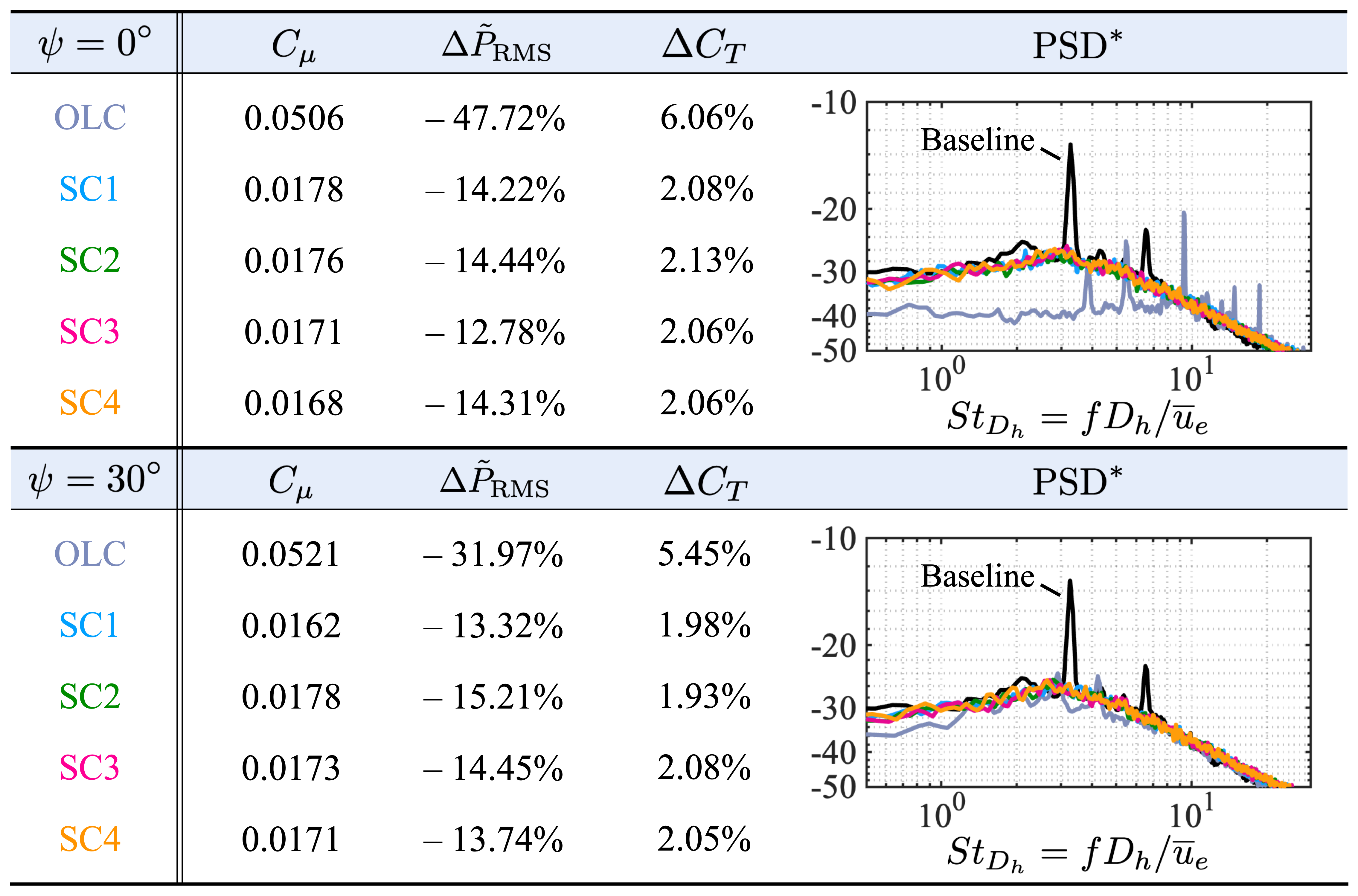}
    \caption{Momentum coefficient, change in surface loading, change in thrust coefficient, and PSD$^*$ for all unconstrained cases.}
    \label{fig:ADFC_unstdns}
\end{figure}

Actuation-induced changes are characterized by the momentum introduced by the controller, surface loading along the aft-deck plate, thrust coefficient, and the pressure spectra through P2, and are shown collectively in Fig.~\ref{fig:ADFC_unstdns} for all adaptive control cases and steady OLC for comparison. 
The $C_\mu$ column shows that employing adaptive control reduces approximately 60\% of the overall momentum introduced to the flow for all feedback control cases compared to steady-blowing actuation. 

To assess the severity of the surface loading, the root-mean-square of the pressure fluctuations along the aft-deck, from the SPTE to the aft-deck trailing edge, is used. 
The change in surface loading is computed as $\Delta \tilde{P}_{\text{RMS}}=\int (\tilde{P}_{\text{RMS},c} - \tilde{P}_{\text{RMS},b})/\tilde{P}_{\text{RMS},b} \times 100$, where $\tilde{P}_{\text{RMS}}=P_{\text{RMS}}/P_\text{ref}$ and the subscripts $c$ and $b$ denote the control case and the baseline condition, respectively.
As shown in the $\Delta \tilde{P}_{\text{RMS}}$ column of Fig.~\ref{fig:ADFC_unstdns}, all adaptive control cases have similar performance in loading reduction, however, steady-blowing actuation demonstrates the greatest reduction. 
This is attributed to the size of the shedding vortices, as steady-blowing results in the consistent formation of small-scale structures.
On the other hand, adaptive control allows some medium-scale structures to form intermittently, as shown in Fig.~\ref{fig:ADFC_flow}d.

The change in thrust associated with each control case is shown in the $\Delta C_T$ column of Fig.~\ref{fig:ADFC_unstdns}, where $\Delta C_T = (C_{T,c}-C_{T,b})/C_{T,b} \times 100$. 
While all cases demonstrate a some improvement in thrust, OLC offers the most enhancement in propulsive power. This is expected due to the steady nature of the control method, which provides a continuous addition of momentum into the flow. On the other hand, adaptive control periodically introduces momentum to the system. The disruption in the vortex formation and consequential reduction in the separation region leads to a modest increase in thrust. For all unconstrained control cases, minimal changes are also observed in the mean flow, as shown in Fig.~\ref{fig:ADFC_flow}a. Since $C_T$ is computed on the mean quantities (Eq.~(\ref{eq:thrust})), this leads to comparable thrust values.

The pressure spectral content through P2 is compared in the $\text{PSD}^*$ column between the baseline, steady OLC, and all feedback control cases to investigate tonal changes in the flow. 
The distributions remain comparable between all adaptive control cases with slight variation in the lower frequency range of $St_{D_h}~<~2.0$. 
The characteristic frequency of $St_{D_h} = 3.28$ is notably missing in all adaptive control cases. 
This is to be expected since the actuator directly interferes with the shedding instability responsible for this tone. 
Interestingly, adaptive control appears to additionally ``smooth out'' other local spikes in the baseline spectrum apart from the resonant tone while preserving its overall profile. 
This further reinforces the ability of adaptive control to target undesired features without disrupting the mean flow.
Meanwhile, steady OLC at $\psi=0\degree$ demonstrates a notable reduction in the broadband content at the expense of newly-introduced high and low frequencies. 

Finally, it is observed that the adaptive control framework is more sensitive to the choice of actuation angle rather than the sensors utilized to inform the snapshot matrices.
This indicates that actuating at an optimal angle is more effective than sensing specific flow features. 
Blowing at either $\psi=0\degree$ or $30\degree$ drives the flow towards comparable states, however, $30\degree$ actuation yields weaker shock patterns.
This indicates that some inclination in the control direction is favorable, as was reported in prior steady control studies~\cite{yeung2024high}. 
As shown in Fig.~\ref{fig:ADFC_unstdns}, the momentum coefficient $C_\mu$, surface loading $\Delta \tilde{P}_{\text{RMS}}$, thrust coefficient $\Delta C_T$, and spectral content are largely comparable across all sensor configurations considered. 
This is promising for practical flow control designs when sensor placement may be limited to physical surfaces.

\subsection{Adaptive Control with Actuator and Model Constraints}

While online DMD demonstrates promising results for real-time adaptive control, the high-speed nature of this flow requires an unrealizable temporal resolution of $St_{D_h}=9.8$, or 98 kHz, to reliably track the dynamics of the dominant shedding instability. 
Therefore, certain constraints are imposed on the model to better represent physical limitations. 
First, the frequency of the actuator is lowered to $St_{D_h}=0.16$, or 1600 Hz.
Second, the system model is updated 10 times prior to computing a new input. 
Third, a subsonic actuation speed of $|u_k/c_{\text{ref}}|\leq0.3$ is also considered. 
Finally, SC4 and $\psi=30\degree$ are chosen for this investigation as it was found that online DMD is not sensitive to the sensor arrangement and SC4 represents the most practical implementation (see Fig.~\ref{fig:AFC_SCs}b). Additionally, $\psi=30\degree$ achieves slightly more reduction in unsteadiness than $0\degree$. 

\subsubsection{Sonic actuation}

\begin{figure}[hbpt]
\centering
    \includegraphics[width=1\textwidth]{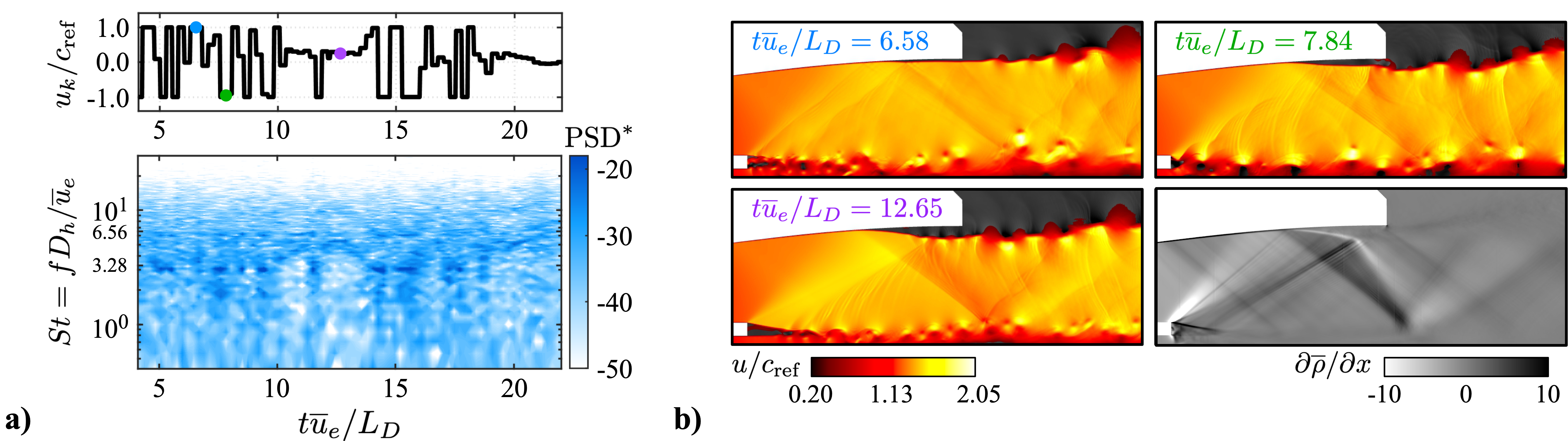}
    \caption{Constrained adaptive control with $|u_k/c_\text{ref}|\leq1.0$. a) Control signal with P2 spectrogram. b) Representative instantaneous and mean flow fields.}
    \label{fig:ADFC_pm1}
\end{figure}

An actuation speed of $|u_k/c_\text{ref}|\leq1.0$ is first discussed and shown in Fig.~\ref{fig:ADFC_pm1}. 
Figure~\ref{fig:ADFC_pm1}a presents the control signal through time along with a spectrogram of the P2 pressure fluctuations, and demonstrates the transient flow behavior in response to the actuation.
The resonant tone $St_{D_h}=3.28$ and its harmonic $St_{D_h}=6.56$ are observed to appear intermittently, specifically during the suction phases of the control signal. During the suction phase, the flow condition returns to the baseline state, as will be discussed later, and is likely the cause for the tone re-emergence.
For $t\overline{u}_e/L_D \in (10,14)$, the resonant tone is mostly eliminated while maintaining a relatively low control input. 
Whenever the tone is detected, the actuator responds accordingly by increasing the input amplitude. 
Although the actuator maintains a low amplitude and appears almost inactive during certain time intervals, such as $t\overline{u}_e/L_D \in (11,14)$ or $(19,22)$, the resonant tone is also absent during these intervals. This suggests that the controller only actuates when necessary, i.e., when the tone is detected, enhancing the overall energy efficiency of the control system.

Figure~\ref{fig:ADFC_pm1}b shows representative streamwise velocity flow fields at the corresponding time instants marked with circles in \ref{fig:ADFC_pm1}a along with a mean streamwise density gradient field. 
When the actuator is in its blowing phase at $t\overline{u}_e/L_D=6.58$, the flow gravitates towards the same flow state as OLC at $\psi=30\degree$ (see Fig.~\ref{fig:ADFC_flow}).
However, during the suction phase at $t\overline{u}_e/L_D=7.84$, the flow appears similar to the baseline case (see Fig.~\ref{fig:base_flow}).
This indicates that blowing pushes the flow to a new state, while suction returns the flow to its original configuration. 
At $t\overline{u}_e/L_D=12.65$, the vortex shedding is notably delayed with a low amplitude forcing, where a temporary stabilization of the shedding instability occurs prior to destabilization.
The primary shock also impinges further downstream along the nozzle and the resultant separation is increased.
Relative to the shedding, the actuation frequency is remarkably lower. This increases the spatial extent of the shock motion, causing the shock train to exhibit a large and irregular breathing pattern. 
The mean flow is also drastically altered, as opposed to feedback control with no constraints. When the control frequency is high, the actuator pulses take place in a much shorter time scale and therefore do not have a significant impact on shock movements.

\subsubsection{Subsonic actuation}

\begin{figure}[hbpt]
\centering
    \includegraphics[width=1\textwidth]{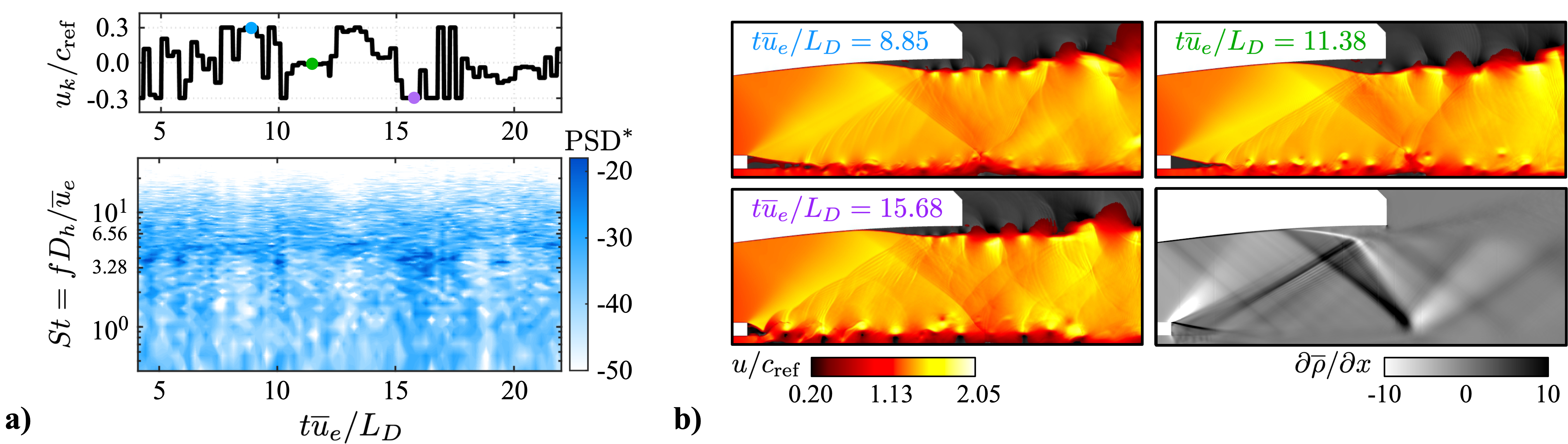}
    \caption{Constrained adaptive control with $|u_k/c_\text{ref}|\leq0.3$. a) Control signal with P2 spectrogram. b) Representative instantaneous and mean flow fields.}
    \label{fig:ADFC_pm03}
\end{figure}

A subsonic actuator is explored, where the actuation speed is constrained to $|u_k/c_\text{ref}|\leq0.3$. 
Figure~\ref{fig:ADFC_pm03}a shows the control signal along with the P2 spectrogram. 
While the dominant tone intermittently vanishes for $|u_k/c_\text{ref}|\leq1.0$, actuating at $|u_k/c_\text{ref}|\leq0.3$ causes the dominant tone to appear sporadically and oscillate between $St_{D_h}=3.28$ and $6.56$. 
Similar to sonic-bounded actuation, the control input adjusts accordingly when the resonant tone is detected. 
Representative streamwise velocity flow fields and a mean streamwise density gradient field are shown in Fig.~\ref{fig:ADFC_pm03}b.
During the blowing phase at $t\overline{u}_e/L_D=8.85$, a temporary stabilization of the shedding instability is achieved. 
As the control amplitude is maintained at $u_k/c_\text{ref}\ge0.0$, the instability quickly destabilizes and generates small-scale vortices, as shown at $t\overline{u}_e/L_D=11.38$, before temporarily stabilizing again. 
This stabilizing and destabilizing behavior is observed to repeat for as long as $u_k/c_\text{ref}\ge0.0$ is maintained, resulting in intermittent vortex shedding. 
When the actuator is in its suction phase at $t\overline{u}_e/L_D=15.68$, the flow again behaves similarly to the baseline configuration. 

While the mean shock structure and location appears similar to control at $|u_k/c_\text{ref}|\leq1.0$, the relative strength of the shocks are stronger for subsonic actuation. 
This is due to the fact that subsonic blowing is insufficient in weakening the primary shock, and only delays the onset of shedding. 
This delay drives the shock forward and induces a larger separation region. 
Consequently, the breathing pattern of the shock train, while still irregular, is comparatively more confined. 
The oscillations occur over a smaller spatial extent, leading to a strengthening of the time-averaged shock. 

\begin{figure}[hbpt]
\centering
    \includegraphics[width=1\textwidth]{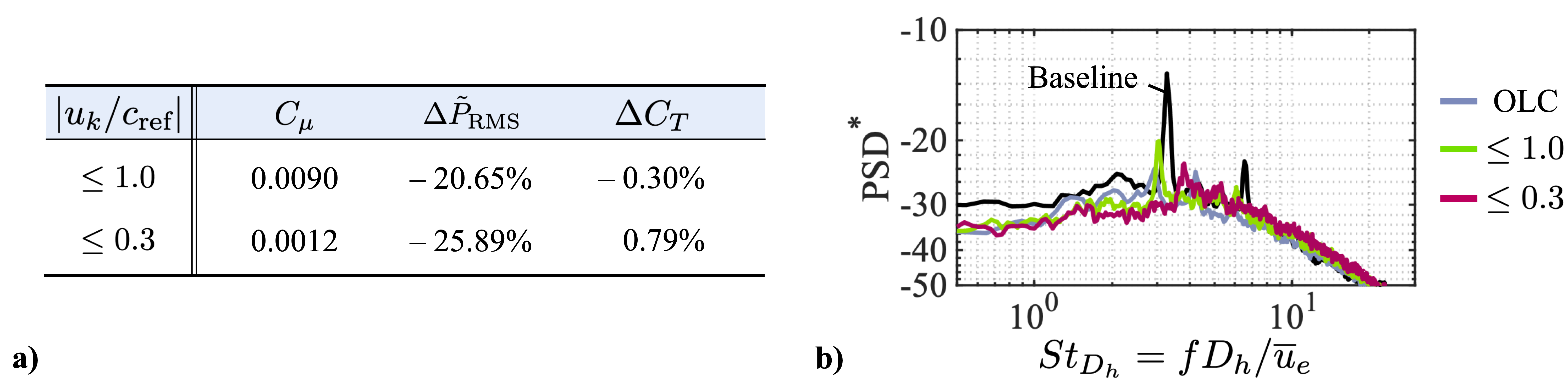}
    \caption{a) Momentum coefficient, change in surface loading, change in thrust coefficient, and b) PSD$^*$ for constrained control cases.}
    \label{fig:ADFC_unstdns_C}
\end{figure}

The momentum coefficient $C_\mu$, change in surface loading $\Delta \tilde{P}_\text{RMS}$, and change in thrust coefficient $\Delta C_T$ are shown in Fig.~\ref{fig:ADFC_unstdns_C}a. 
Enforcing restrictions on the feedback control model yields less momentum introduced to the flow compared to the unrestricted cases (see Fig.~\ref{fig:ADFC_unstdns}).
As surface loading is directly related to shedding, low-frequency actuation generates more small-scale structures and achieves greater reduction in surface loading. 
Moreover, the repeated stabilization and destabilization of the shear layer instability under subsonic actuation leads to stronger vortex suppression compared to sonic-bounded actuation, giving higher reduction in surface loading. 
The thrust generated remains comparable to the baseline case, with the unconstrained control methods achieving higher propulsion (see Fig.~\ref{fig:ADFC_unstdns}).
The PSD$^*$ of the control signals ($u_k/c_\text{ref}$ in Fig.~\ref{fig:ADFC_pm1}a and \ref{fig:ADFC_pm03}a) reveal an effective frequency of $St_{D_h}=0.031$ and $0.008$, respectively.
Lastly, the PSD$^*$ of the P2 pressure signal, shown in Fig.~\ref{fig:ADFC_unstdns_C}b, demonstrates a notable reduction in the resonant tone, however, its suppression is not as effective as unconstrained adaptive control. 

\subsection{Tone Contribution of Intermittent Events and Statistical Analysis}
\label{sec:events}

\begin{figure}[hbpt]
\centering
    \includegraphics[width=1\textwidth]{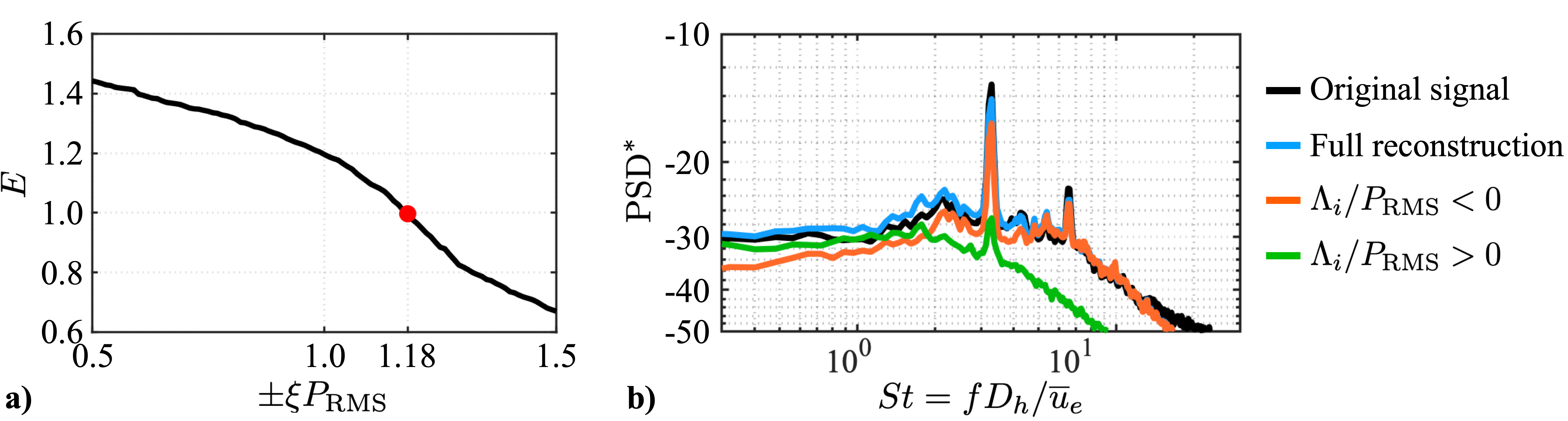}
    \caption{a) Optimal $\xi$ for signal reconstruction. b) PSD$^*$ of the original and reconstructed signals.}
    \label{fig:event_swp_PSD}
\end{figure}

To investigate the contribution of intermittent events to the dominant tone, the pressure time series through P2 is reconstructed, as shown in Fig.~\ref{fig:event_method}a. 
In this region, it is found that the acoustic Mach number is approximately $\text{M}_a\approx0.6$, making the splitter plate shear layer acoustically subsonic. 
The choice of $\xi$ (see Section~\ref{sec:event_method}) is decided such that the energy ratio, defined as $E=\text{var}(\phi)/\text{var}(P/P_\text{RMS})$, approaches unity. 
A value of $E>1$ indicates the reconstruction has added energy into the signal, while $E<1$ indicates an underestimation of the energy. 
It is found that $\xi=1.18$ achieves a near-unity reconstruction of $E=0.997$, as shown in Fig.~\ref{fig:event_swp_PSD}a. It is noted that $\xi=1.18$ is unique to P2 and $\xi$ is dependent on the signal location and flow condition~\cite{kearney2013intermittent}.  

A power spectral density of the original and reconstructed signal for the baseline case is presented in Fig.~\ref{fig:event_swp_PSD}b. The reconstructed is performed in three ways. The full reconstruction is carried out as described in Section~\ref{sec:event_method}. The cases with $\Lambda_i/P_\text{PRMS} < 0$ and $\Lambda_i/P_\text{PRMS} > 0$ correspond to reconstructions based solely only the negative-valued or positive-valued events, respectively. 
It is observed that both the original and full reconstruction spectra overlap, and the resonant tone and overall behavior are successfully reproduced by the full reconstruction.  
As shown in \ref{sec:apdx_event}, the slight underestimation of the amplitude is attributed to the event definition of the reconstructed signal, where some tone-contributing regions of the original signal are not captured. 
Despite this, the events alone account for approximately 80\% of the amplitude.
This suggests that the intermittent events are a primary contributor to the dominant frequency. 
The decrease in the high-frequency content for the reconstruction is also expected since the model function has filtered out the high-frequency behavior in the original signal. 
A further decomposition into the negative- and positive-valued components reveal contributions made by low- and high-pressure events, respectively.
The PSD$^*$ demonstrates that the low-pressure events contribute greatly to the resonant tone, whereas the high-pressure events predominantly contribute to the low-frequency broadband content.

\begin{figure}[hbpt]
\centering
    \includegraphics[width=1\textwidth]{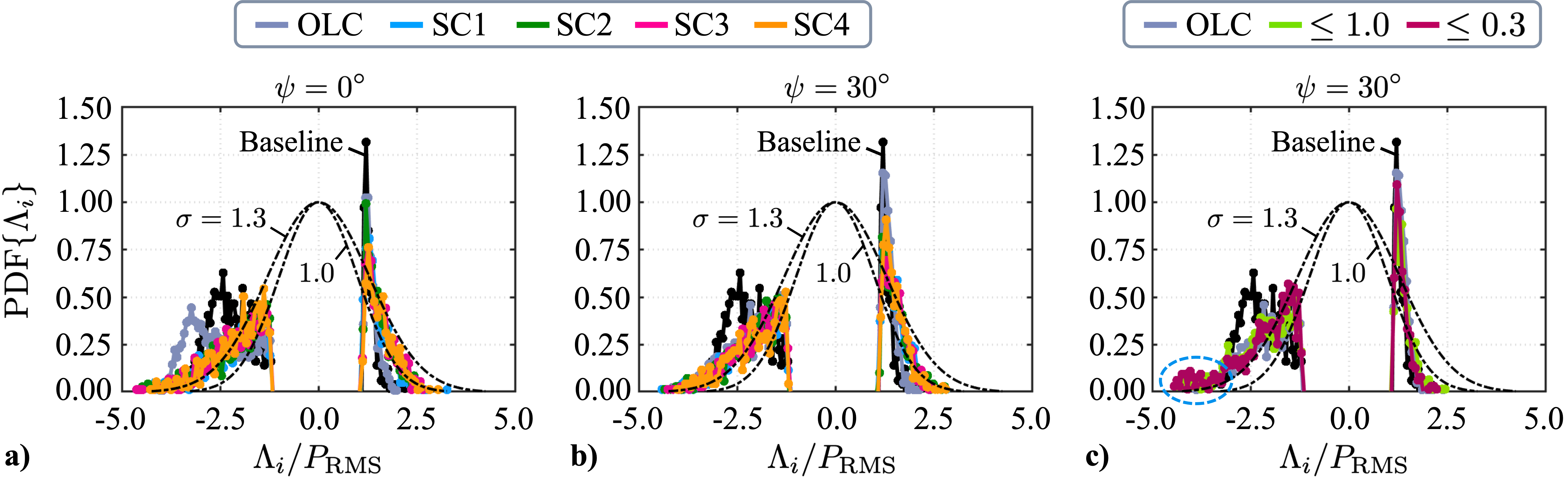}
    \caption{PDF of the event amplitudes $\Lambda_i$ for unconstrained control with a) $\psi=0\degree$, b) $\psi=30\degree$, and c) constrained control.}
    \label{fig:event_PDFs}
\end{figure}

The probability density function (PDF) of the event amplitudes $\Lambda_i/P_\text{RMS}$ is presented for all cases in Fig.~\ref{fig:event_PDFs}. The PDFs reveal how frequently the intermittent events occur and their corresponding strength, and the sharp cut-off at $\pm1.18~P_\text{RMS}$ is a consequence of the event definition. 
Two Gaussian distributions $G(x)=\text{exp}(-0.5(x/\sigma)^2)$ are also indicated for reference with standard deviations of $\sigma=1.0$ and 1.3.
The baseline PDF distribution is observed to be highly skewed, with the intermittent events containing strong bursts of energy that are correlated to the resonant tone. 
All unconstrained adaptive control cases for $\psi=0\degree$ and $30\degree$ are shown in Fig.~\ref{fig:event_PDFs}a and b respectively, along with the steady OLC for comparison.
For $\psi=0\degree$, the PDFs demonstrate that all unconstrained cases can effectively suppress the low-pressure events, although suppression of high-pressure events varies by case. 
These PDFs better resemble a normal distribution, albeit the PDF is not perfectly normal and is slightly skewed left. 
The result of the low-pressure event suppression is directly reflected in the corresponding PSD$^*$ spectra shown in Fig.~\ref{fig:ADFC_unstdns}, where the resonant tone has been eliminated. 
The broadband content is still preserved in these cases, as the high-pressure events have not been entirely mitigated.
On the other hand, OLC at $\psi=0\degree$ amplifies the low-pressure events while damping high-pressure events. This results in the production of higher tones, as shown in Fig.~\ref{fig:ADFC_unstdns}, but with a reduction in overall broadband content. 
For $\psi=30\degree$, all control cases including OLC effectively suppress the low-pressure events, while high-pressure event attenuation varies by case. As a result, the resonant tone is removed while preserving the overall PSD$^*$ profile, as shown in Fig.~\ref{fig:ADFC_unstdns}.

Finally, the PDFs for constrained adaptive control at $|u_k/c_\text{ref}|\leq1.0$ and $\leq0.3$ are shown in Fig.~\ref{fig:event_PDFs}c. Similar to the unconstrained cases, the high-pressure events are not fully mitigated. Although the PDF distribution appears similar to that of the unconstrained cases, there is a slight amplification in the strength of the low-pressure events, as indicated with a dashed blue circle. Consequently, the tone for these cases is shifted to a new frequency, albeit at a lower amplitude, as shown in Fig.~\ref{fig:ADFC_unstdns_C}b.

\section{Conclusion}
\label{sec:conc}

An adaptive control framework is applied to a supersonic dual-stream jet flow using two-dimensional direct numerical simulations. 
The jet flow exhibits various interconnected and adverse phenomena, such as vortex shedding, shock systems, shock-boundary-layer interactions, and shock-induced separation. A high-amplitude tone that permeates the near- and far-fields is found to originate from the mixing of the two primary flows within the nozzle. 
Actuation is introduced along the vertical surface of the splitter plate trailing edge.
Online dynamic mode decomposition is employed to approximate and update the underlying system dynamics using limited sensor measurements. Various sensor configurations are explored, where each measures the instantaneous pressure fluctuations. It is found that online DMD-based control is not sensitive to the sensor placements, but rather the actuation angle.  

Adaptive feedback control with an unconstrained controller frequency at sonic-bounded amplitudes is shown to break down the large-scale coherent structures with little disturbance to the mean flow. Moreover, approximately 60\% less momentum is introduced to the flow when compared to steady-blowing open-loop control. 
Spectral analysis verifies the elimination of the high-frequency tone, and the overall profile of the pressure spectra are observed to be ``smoothed out" of any local spikes. 
Adaptive control using online DMD effectively targets the instability while preserving the mean flow. 

As high-speed flows may require unfeasible temporal resolutions to reliably track certain flow dynamics, additional constraints are placed on the model to represent physical limitations, and both sonic- and subsonic-bounded actuation is considered. 
It is found that the breathing pattern of the shock train becomes highly irregular, with sonic-bounded actuation exhibiting a larger spatial extent in shock motion.
Meanwhile for subsonic actuation, the shear layer instability undergoes repeated, yet irregular, stabilizing and destabilizing events. 
This leads to a greater suppression in vortex generation and consequently the surface loading. 

Extracting information regarding intermittent events from a local pressure signal reveals that such events are a large contributor to the dominant tone. Further decomposition into low- and high-pressure events demonstrates that the low-pressure events constitutes a majority of the tone, while high-pressure events make up the low-frequency broadband content. 
The unconstrained adaptive controller effectively targets and suppresses low-pressure events, leading to the elimination of the resonant tone. Meanwhile, the high-pressure events are not entirely mitigated, resulting in the preservation of the overall broadband content. 
On the other hand, the constrained feedback controller has a slight amplification of low-pressure events, although they remain largely suppressed overall. This results in a partial attenuation of the resonant tone. 

Future investigation into employing adaptive control at more experimentally practical locations, such as along the aft-deck, could prove valuable. Furthermore, exploring the use of two or more actuators to target both the upper shear layer and splitter plate shear layer may offer significant benefits.

\appendix
\renewcommand{\thesection}{Appendix~\Alph{section}}

\section{Linear Time-Varying Spring-Mass-Damper System}
\label{sec:apdx_SMD}

A spring-mass-damper problem is utilized to investigate the robustness of online DMD in accurately modeling the dynamics of a time-varying system under the influence of external forcing $\mathcal{F}(t)$.
This canonical problem helps demonstrate the behavior of online DMD, and is represented as
\begin{equation}
    \gamma\ddot{x} + \beta\dot{x} + \kappa x = \mathcal{F}(t).
\end{equation}
Rewriting in state-space form, this becomes 
\begin{equation}
    \bm{\dot{x}} = \mathcal{A}\bm{x} + \mathcal{B}\bm{u}
    , ~~~~~~~~~~
    \mathcal{A} = \begin{bmatrix}
                    0               & 1        \\
                    -\kappa/\gamma & -\beta/\gamma \\ 
                    \end{bmatrix}
    , ~~~~~~~~~~
    \mathcal{B} = \begin{bmatrix}
                    0      \\
                    1/\gamma \\ 
                    \end{bmatrix}.
\label{eq:SMD}
\end{equation}
Here, $\gamma$ is defined as $\gamma = 1 + \epsilon t$ and $u = \sin(\omega t)$, and thus represents a spring-mass-damper system with time-varying mass. 
The parameters of the system are set to $\kappa = 8$, $\beta = 1$, $\omega = 1$, and $\epsilon = 0.5$. 
Since this exercise only serves to investigate system identification using online DMD rather than simulating the effects of control, the signal of $u(t)$ is preserved in its sinusoidal form and never externally modulated.
Finally, the initial conditions and the weighting factor are designated to be $\bm{x}_0 = [0, 1]^T$ and $\Omega = 0.3$, respectively.

\begin{figure}[hbpt]
    \centering
    \includegraphics[width=1\textwidth]{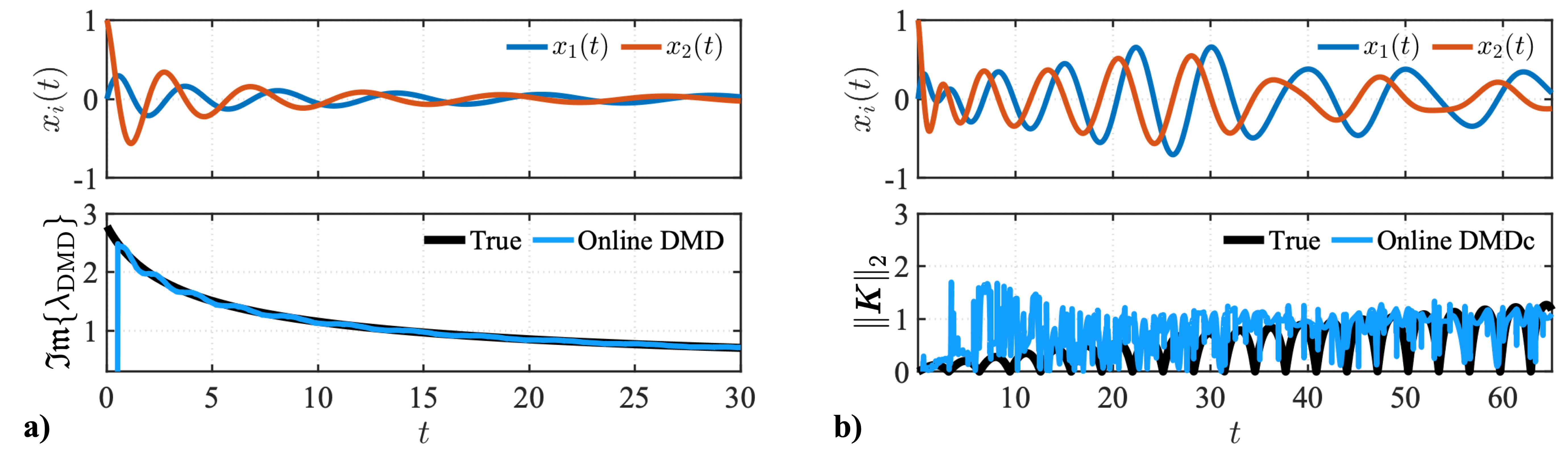}
    \caption{State evolution through time and online DMD estimation of a) $\dot{\bm{x}} = \mathcal{A}\bm{x}$ and b) $\dot{\bm{x}} = \mathcal{A}\bm{x} + \mathcal{B}\bm{u}$.}
    \label{fig:SMD}
\end{figure}

First, online DMD is applied to the above system without the influence of forcing, i.e., $\dot{\bm{x}} = \mathcal{A}\bm{x}$, as shown in Fig.~\ref{fig:SMD}a.
It is evident that the system holds a time-varying frequency, which may be represented through the imaginary component of the DMD eigenvalues. 
It is noted that DMD returns the discrete-time eigenvalues $\mu_{\text{DMD}}$ calculated from the weighted snapshots, which are related to the continuous-time DMD eigenvalues $\lambda_{\text{DMD}}$ as $\mu_{\text{DMD}} = e^{\lambda_{\text{DMD}} \Delta t}$~\cite{zhang2019online}.
The online DMD eigenvalues closely follow the true eigenvalues of the system, while the discrepancy near $t=0$ is attributed to the choice of initialization. Here, initialization was defined as $\bm{A}_0 = \textit{\textbf{0}}$ and $\bm{P}_0 = \alpha \bm{I}$, which reflects how the algorithm is implemented into the numerical fluid flow solver. This initial discrepancy may be removed by choosing $\bm{A}_0 = \bm{Y}_w \bm{X}_w^\dag$ and $\bm{P}_0 = (\bm{X}_w \bm{X}_w^T)^{-1}$.

The algorithm is then applied to the system with external forcing, i.e., $\dot{\bm{x}} = \mathcal{A}\bm{x} + \mathcal{B}\bm{u}$, as shown in Fig.~\ref{fig:SMD}b.
As the continuous-time eigenvalues can no longer be obtained with the addition of forcing due to the dynamical matrix becoming rectangular, the feedback gain matrix $\bm{K}$ is used to measure the performance of online DMD. 
The $\| \cdot \|_2$ norm of the gain $\bm{K}$ is presented, where the true feedback gain matrix $\bm{K}_{\text{true}}$ is compared to  $\bm{K}_{\text{DMD}}$ estimated from online DMD. 
In the $\bm{K}_{\text{true}}$ calculation, the exact continuous-time $\mathcal{A}$ and $\mathcal{B}$ matrices from Eq.~(\ref{eq:SMD}) are input into MATLAB's built-in function \textit{lqr}, which calculates the optimal gain matrix based on the continuous-time algebraic Riccati equation. On the other hand, the DMD-estimated gain matrix $\bm{K}_\text{DMD}$ is computed by using the estimated discrete-time matrices $\bm{A}_k$ and $\bm{B}_k$ at each time instant $t_k$ and iterating the discrete-time algebraic Riccati equation (Eqs.~(\ref{eq:DARE},\ref{eq:DARE-K})). It is observed that accuracy is lost between the true and estimated gains.
However after some time, $\bm{K}_\text{DMD}$ begins to track with $\bm{K}_{\text{true}}$, suggesting the dynamical models approximated from online DMD require an initialization interval to stabilize before beginning to reflect the true dynamics.

\section{Event Definition and Resonant Tone Contributions}
\label{sec:apdx_event}

\begin{figure}[hbpt]
    \centering
    \includegraphics[width=1\textwidth]{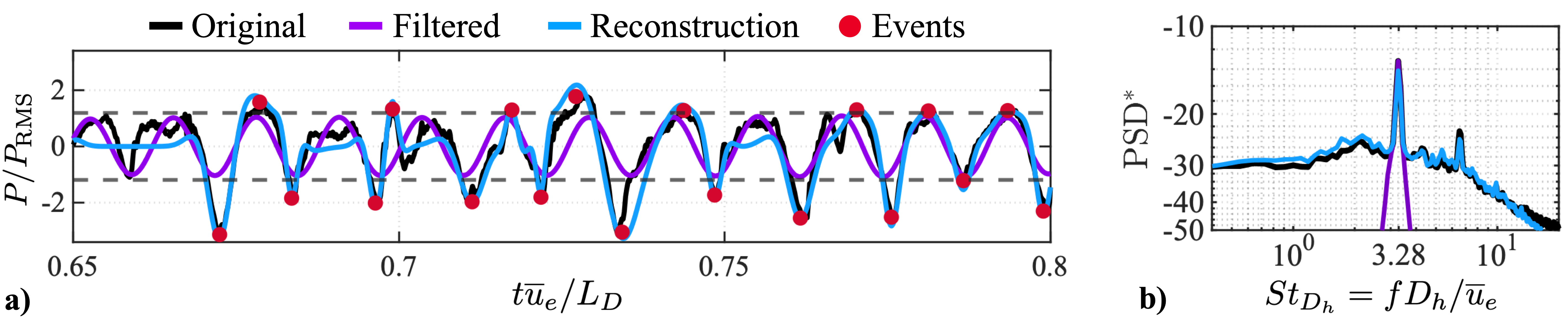}
    \caption{a) Original, filtered, and reconstructed signals. b) Corresponding PSD$^*$ spectra.}
    \label{fig:filt_recon}
\end{figure}

The pressure time series through P2 is filtered using a zero-phase bandpass infinite impulse response (IIR) filter to isolate a narrow frequency band centered on the resonant tone. Figure~\ref{fig:filt_recon}a compares the original, filtered, and reconstructed signals.
The reconstruction captures the general amplitude trends of the original signal, however, occasionally over- and underestimates the filtered signal. 
Moreover, some regions in the narrowband signal are excluded in the event-based reconstruction.
This leads to a slight underestimation of the peak amplitude in the corresponding PSD$^*$ spectra, as shown in Fig.~\ref{fig:filt_recon}b. 
Nonetheless, the amplitude difference is small and the events capture a dominant contribution to the resonant tone. 

\section*{Funding Sources}

This material is based upon work supported by the Air Force Office of Scientific Research under award number FA9550-23-1-0019 (Program Officer: Dr.~Gregg Abate) as well as the National Science Foundation Graduate Research Fellowship under Grant no. 2139766.

\section*{Acknowledgments}
We gratefully acknowledge the Extreme Science and Engineering Discovery Environment (XSEDE), now part of ACCESS, supported by National Science Foundation grant number PHY250051, and the computational resources provided by Purdue Anvil. We also thank Dr. Mark N. Glauser, Dr. Datta V. Gaitonde, and Dr. Fernando Zigunov for the insightful discussions.

\bibliography{bibliography}

\end{document}